\documentclass[11pt]{article}

\pdfoutput=1

\usepackage[dvipsnames]{xcolor}
\usepackage{bm}
\usepackage[top=80pt,bottom=85pt,left=85pt,right=85pt]{geometry}
\usepackage{amsmath,amssymb,bbm,graphicx,float,color,tikz,cite,savesym,wasysym,slashed}

\usepackage{caption}
\usepackage{subcaption}
\usepackage{bigints}
\usepackage[debug,pageanchor=false]{hyperref}
\definecolor{link}{rgb}{.8,.15,.1}
\hypersetup{colorlinks=true,linkcolor=link,citecolor=link,urlcolor=link,linktocpage}

\renewcommand{\Re}{\mathrm{Re}}    
\renewcommand{\Im}{\mathrm{Im}}  
\newcommand{\del}{\partial}
\newcommand{\ii}{\mathrm{i}}
\newcommand{\dd}{\mathrm{d}}
\newcommand{\ee}{\mathrm{e}}

\newcommand{\beq}{\begin{equation}}
\newcommand{\eeq}{\end{equation}}

\newcommand{\cC}{{\mathcal C}}
\newcommand{\cD}{{\mathcal D}}
\newcommand{\cE}{{\mathcal E}}

\newcommand{\calf}{{\mathcal F}}
\newcommand{\calg}{{\mathcal G}}
\newcommand{\calh}{{\mathcal H}}
\newcommand{\calm}{{\mathcal M}}
\newcommand{\calv}{{\mathcal V}}
\newcommand{\calk}{{\mathcal K}}
\renewcommand{\d}{\mathrm{d}}

\usepackage[utf8]{inputenc}
\usepackage{uniinput,verbatim}
\newcommand{\ie}{\textit{i.e.,}~}

\makeatletter
\newcommand{\raisemath}[1]{\mathpalette{\raisem@th{#1}}}
\newcommand{\raisem@th}[3]{\raisebox{#1}{$#2#3$}}
\makeatother

\usepackage{relsize}
\newcommand{\s}[1]{{\mathsmaller{#1}}}
\newcommand{\lp}{\ell_\textrm{\tiny P}}

\usepackage{tabto}

\usepackage[low-sup]{subdepth}

\def\d {{\rm d}}
\def\del          {\partial}

\def\ii           {{\rm i}}

\def\Re           {{\rm Re\hskip0.1em}}
\def\Im           {{\rm Im\hskip0.1em}}

\def\cala         {{\cal A}}

\def\calc         {{\cal C}}
\def\cald         {{\cal D}}
\def\cale         {{\cal E}}
\def\calf         {{\cal F}}
\def\calg         {{\cal G}}

\def\calh         {{\cal H}}

\def\calk         {{\cal K}}

\def\calm         {{\cal M}}
\def\caln         {{\cal N}}
\def\calo         {{\cal O}}
\def\calp         {{\cal P}}

\def\calv         {{\cal V}}

\makeatletter
\@addtoreset{equation}{section}
\makeatother

\begin{document}

	\begin{titlepage}

	\begin{center}

	\vskip .5in %.3in 
	\noindent

	{\Large \bf{On the Stability of String Theory Vacua}}

	\bigskip\medskip

	 Suvendu Giri,$^{1,2}$ Luca Martucci,$^{3}$ Alessandro Tomasiello$^{2,4}$\\

	\bigskip\medskip
	{\small 
$^1$ Dipartimento di Fisica, Universit\`a di Milano-Bicocca, \\ Piazza della Scienza 3, I-20126 Milano, Italy
\\	
	\vspace{.3cm}%{.1cm}
$^2$ INFN, sezione di Milano-Bicocca, Piazza della Scienza 3, I-20126 Milano, Italy
\\	
	\vspace{.3cm}%{.1cm}
$^3$ Dipartimento di Fisica e Astronomia, Universit\`a degli Studi di Padova\\
\& INFN sezione di Padova, Via F. Marzolo 8, I-35131 Padova, Italy
\\	
	\vspace{.3cm}%{.1cm}
$^4$ Dipartimento di Matematica, Universit\`a di Milano-Bicocca, \\ Via Cozzi 55, I-20126 Milano, Italy	
		}

	\vskip .5cm %.3cm
	{\small \tt suvendu.giri@unimib.it, luca.martucci@pd.infn.it, alessandro.tomasiello@unimib.it}
	\vskip .9cm %.6cm
	     	{\bf Abstract }
	\vskip .1in
	\end{center}

	\noindent

\noindent Vacuum compactifications may suffer from instabilities under small perturbations or tunnel effects; both are difficult to analyze. In this paper we consider the issue from a higher-dimensional perspective. We first look at how stability works for supersymmetric vacua, where it is widely expected to hold. We first show that the nucleation of brane bubbles in type II AdS compactifications is forbidden in the probe approximation by a simple argument involving pure spinors and calibrations. We then adapt familiar positive-energy theorems directly to M-theory and type II supergravity, rather than to their effective lower-dimensional reductions, also showing how to consistently include localized sources. We finally initiate an analysis of how these arguments might be extended to non-supersymmetric vacua. In M-theory, at the lower-derivative level, we find that the most natural modifications fail to stabilize the skew-whiffed and Englert vacua.

	\noindent

	\vfill
	\eject

	\end{titlepage}

\tableofcontents

\section{Introduction and summary} % (fold)
\label{sec:intro}

In a gravitational theory, spacetime becomes dynamical. This creates the possibility that it might change drastically, perhaps with a catastrophic change in vacuum energy. For the Minkowski vacuum of general relativity, such worries were put to rest long ago by mathematical theorems \cite{schoen-yau,witten-pe}. If matter satisfies the dominant energy condition, an appropriately defined total energy, computed in terms of the boundary behavior of the metric, is non-negative, and can only be zero for Minkowski space itself; so not even a quantum tunneling event can trigger spacetime decay. 

This result was later generalized in various directions \cite{Abbott:1981ff,gibbons-hull-warner,hull-positivity,boucher}. One of the lessons was that gravity can help stabilize scalar potentials that in field theory would lead to vacuum decay, confirming considerations from small fluctuations \cite{breitenlohner-freedman} and instantons \cite{coleman-deluccia}. Another is that supersymmetry often plays a crucial role, ultimately because the Hamiltonian is the square of the supercharges.  (Indeed even the argument in \cite{witten-pe} for pure gravity relies on introducing auxiliary spinors which are covariantly constant at infinity, inspired by the vacuum supersymmetric spinors for the simplest supergravity models.) In particular, it was shown in various $d=4$ examples that supersymmetric vacua are stable even if the model does not satisfy the dominant energy condition \cite{gibbons-hull-warner}.

In theories with extra dimensions such as string theory, however, new decay channels open up. This is demonstrated in a spectacular fashion by the $\mathrm{Mink}_4\times S^1$ Kaluza--Klein (KK) vacuum of pure five-dimensional gravity, which can decay via the nucleation of a bubble of different topology, created by recombining internal and external directions; from the four-dimensional perspective this appears as a \emph{bubble of nothing} where spacetime ends \cite{witten-bubble-nothing}. (The supersymmetry-inspired argument in \cite{witten-pe} does not apply, because on the new topology there exist no spinors that are asymptotically covariantly constant.) In string theory, many other similar decay channels are provided by brane bubbles, where again spacetime can end or across which the flux quanta can change (with early examples in \cite{horowitz-orgera-polchinski,dowker-gauntlett-gibbons-horowitz-nucleation,maldacena-michelson-strominger}). While an effective four-dimensional approach might go a long way towards establishing vacuum stability, it cannot capture such processes, which involve the dynamics of higher KK modes.\footnote{Effective theories of compactifications often not only violate the dominant energy condition, but naively seem to have a scalar potential that is unbounded from below, coming from the internal integral of the scalar curvature \cite{Hertog:2003ru}. This issue was however resolved in \cite{douglas-warping}.} This issue is made all the more pressing by the absence of a scale separation between the KK scale and the vacuum energy in many AdS vacua. While in some theories there was recent progress (beginning with \cite{malek-samtleben-kk}) in computing the full infinite KK mass spectrum, this still only covers small fluctuations.

In this paper, we attack this problem directly in ten or eleven dimensions. For supersymmetric $\mathrm{Mink}_4 \times M_d$ or $\mathrm{AdS}_4 \times M_d$ vacua, $d=6,7$, stability is of course widely expected for the reasons we recalled earlier; but quite surprisingly, it was never explicitly shown before. Our original motivation came from vacua that break supersymmetry: de Sitter vacua are expected to be metastable at best, while anti-de Sitter (AdS) vacua have been conjectured to be either unstable or metastable by a generalization of the weak gravity conjecture \cite{ooguri-vafa-ads,freivogel-kleban} -- see for instance \cite{danielsson-dibitetto-vargas-dw,danielsson-dibitetto-vargas-swamp,ooguri-spodyneiko,apruzzi-deluca-gnecchi-lomonaco-t,bena-pilch-warner,garciaetxebarria-montero-sousa-valenzuela,Guarino:2020jwv,apruzzi-deluca-lomonaco-uhlemann,Bomans:2021ara,Marchesano:2021ycx,suh-ads4} for a sample of recent papers discussing  non-perturbative (in)stability in string theory.

In the supersymmetric case, as an appetizer we first consider brane bubbles in the probe approximation. Such processes have been shown to destabilize various non-supersymmetric vacua, beginning with \cite{dowker-gauntlett-gibbons-horowitz-nucleation,maldacena-michelson-strominger}; we show why these can never happen in the supersymmetric case, with a simple argument involving the type II pure spinor equations \cite{gmpt2}. 

We then attack the problem in full generality by adapting the strategy of the positive energy theorem \cite{witten-pe,nester,Abbott:1981ff,gibbons-hull-warner,hull-positivity,boucher} to compactifications. We consider a spinor $\epsilon$ that is \emph{asymptotically supersymmetric}: namely, it is asymptotic to a supersymmetric spinor of the vacuum. From this  we define a notion of \emph{BPS energy} $I(\epsilon)$ by integrating the Hodge-dual of a certain two-form $E_2$, bilinear in $\epsilon$, on the $S^2\times M_d= \partial S$ asymptotic boundary of a spatial slice $S$. Using Stokes' theorem, $I(\epsilon)= \int_S \d *\! E_2$. Our main result is a formula for the divergence $*\,\d *\! E_2$, which we obtained both for eleven-dimensional and type II supergravity. In the former case, for example, we show that on-shell and in absence of branes the divergence satisfies the simple identity $\nabla_\s{M} E^{\s{MN}} = \overline{{\mathcal D}_\s{M} \epsilon} \Gamma^{\s{MPN}} {\mathcal D}_\s{P} \epsilon$,
where $\delta \psi_\s{P}= {\mathcal D}_\s{P} \epsilon$ is the gravitino supersymmetry transformation; several flux and curvature terms conspire on the right-hand side to reconstruct the bulk equations of motion, which we can then set to zero. (The type II result is similar, but also contains the operator ${\mathcal O}$ in the dilatino transformations, $\delta \lambda = {\mathcal O} \epsilon$.) 

The divergence $\nabla_\s{M} E^{\s{MN}}$ has the same form as in $d=4$, so the remaining steps are standard. Namely, let $\epsilon$ be asymptotically supersymmetric and such that the \emph{Witten condition} $\Gamma^a {\mathcal D}_a \epsilon=0$ holds, where $a$ is a flat index tangent to $S$. A gamma matrix identity  now implies $I(\epsilon)= \int_S ({\mathcal D}^a\epsilon)^\dagger({\mathcal D}_a\epsilon)\ge 0$, which is non-negative and only vanishes for supersymmetric vacua, as required. 

In presence of branes, the bulk equations of motion produce delta-like contributions to $I(\epsilon)$; using supersymmetry and the theory of calibrations as generalized in \cite{Martucci:2011dn}, we show that these contributions are still all positive. (This is similar in  spirit to how fields in multiplets not containing the graviton give extra contributions to $\nabla_\mu E^{\mu \nu}$ in $d=4$ supersymmetric models \cite{gibbons-hull-warner}.) 

This proves stability in the supersymmetric case, up to a couple of technical details which are expected to hold and were checked in similar situations in existing literature. First, for AdS vacua one needs to check that $I(\epsilon)$ is independent of the boundary of the spatial slice $\partial S$; or in other words, that the BPS energy is conserved. In $d=4$ this can be done for example by relating it to the energy defined in the covariant phase space formalism, which is conserved \cite{Hollands:2005wt}. Second, one needs to prove that an $\epsilon$ that is asymptotically supersymmetric and satisfies the Witten condition $\Gamma^a {\mathcal D}_a \epsilon=0$ always exists. Here we check a standard formal argument; a rigorous analytical proof was given in \cite{Parker:1981uy} for the $d=4$ case (with an AdS$_4$ generalization in \cite{amsel-hertog-hollands-marolf}), and in \cite{Dai:2003jr,Dai:2004yd} for Calabi--Yau compactifications.

Having worked out the argument for stability of supersymmetric vacua, we finally proceed to non-supersymmetric ones. The idea is now to look for an operator ${\mathcal D}'_\s{M}$ different from the one appearing in supersymmetry, trying to i) solve ${\mathcal D}'_\s{M} \epsilon= 0$ and ii) to make sure that the positivity argument still holds. In $d=4$ models \cite{boucher} this leads to the idea of \emph{fake superpotential}: a $W$ which is related to the scalar potential by the usual supergravity formula $V= 2 (\partial_\phi W)^2 - 3 W^2$, but which isn't the supersymmetric one. There is no direct analogue of this in our higher-dimensional setting; in M-theory, the most general gauge-invariant operator  ${\mathcal D}'_\s{M}$ at one-derivative level is obtained by changing the coefficients of the flux terms in ${\mathcal D}_\s{M}$, and adding a new term proportional to $\Gamma_\s{M}$. 

We are able to satisfy condition i) above on some simple AdS$_4$ vacua: the \emph{skew-whiffed} ones \cite{duff-nilsson-pope-skew,duff-nilsson-pope}, obtained from supersymmetric ones by reversing the sign of $G$, and the \emph{Englert} vacua \cite{englert-susybr}, where internal flux components are added. In both of these cases we find multiple violations to the stability argument. One might want to conclude from this absence of protection that these vacua are in fact unstable. Of course it is possible, however, that a more sophisticated modification of the supersymmetric argument does work for them, or that a completely different protection mechanism is at play. We hope our methods can be refined in the future to settle the issue one way or another.

In section \ref{sec:bubbles} we warm up, as we mentioned, by giving an argument that forbids brane bubbles in the probe approximation for AdS supersymmetric vacua. In section \ref{sec:rev} we switch gears and review positive-energy theorems and their applications to stability in $d=4$ theories. This provides a blueprint for subsequent developments: in section \ref{sec:m} we show positivity for M-theory, and in section \ref{sec:ii} for type II supergravity. Finally we consider supersymmetry-breaking in section \ref{sec:susy-br}, mainly in M-theory. Many technical aspects of the computations are provided in detail in the appendices.

% section intro (end)

\section{Probe brane bubbles} % (fold)
\label{sec:bubbles}

A vacuum may be unstable under tunneling effects. The probability of such a decay can be computed in Euclidean signature, by looking for bubbles that connect the old vacuum at infinity with a new one at their core. 
In this section we consider the simple case of decays mediated by branes in AdS vacua, in the \textit{probe approximation}, namely disregarding their back-reaction. This is of course not always appropriate, but we consider it here as a warm-up: focusing for concreteness on D-brane bubbles in type II AdS$_4$ vacua, we are able to give a very simple argument that such bubbles cannot be nucleated. The argument is generalizable to other settings and  directly exploits the ten-dimensional geometrical structure of the background. In other words, it does not rely on any effective four-dimensional description, which may be problematic in absence of scale separation. 

\subsection{Preliminary general remarks} % (fold)
\label{sub:inst}

We start by reviewing  AdS$_d$ tunneling mediated by a probe $(d-2)$-brane charged under a $d$-form field strength $F_d=\d A_{d-1}$.\footnote{In the simplest $d$-dimensional setting in which only the metric and $A_{d-1}$ appear as dynamical bulk fields, it is relatively easy to take into account backreaction effects  \cite{brown-teitelboim}. This is much harder for generic string/M-theory AdS vacua, in particular in absence of scale separation.} We first set up the formalism in a $d$-dimensional theory, and then discuss the modifications in the presence of extra dimensions.

In global coordinates, the Euclidean AdS$_d$ metric is
\begin{equation}\label{eq:glo-eads}
	\dd s^2_{\mathrm{EAdS}_d}= L^2(\dd r^2 + \sinh^2 r\, \dd s^2_{S^{d-1}})\,,
\end{equation}
where $L$ is the AdS radius.
This metric has an $\mathrm{SO}(1,d)$ isometry group. As in \cite{coleman-deluccia,Abbott:1985kr}, the dominant  instantonic bubbles are expected to be invariant under the largest possible amount of symmetries. Hence, without loss of generality we can focus on Euclidean bubbles wrapping the $S^{d-1}$ in (\ref{eq:glo-eads}), which are invariant under the  isometry subgroup $\mathrm{SO}(d)$.\footnote{The bubbles considered in \cite{maldacena-michelson-strominger} would superficially not look spherically symmetric, but in fact they are; see \cite[Sec.~4.1.2]{apruzzi-deluca-gnecchi-lomonaco-t}.}

The semiclassical decay rate is controlled by a $(d-1)$-dimensional  Euclidean action of the form
\begin{equation}\label{eq:S-brane}
	S=  \tau \int \dd^{d-1} \sigma \sqrt{-g} -  q   \int A_{d-1}\,,
\end{equation}
with some $\tau$ and $ q  $  representing the tension and charge of the brane. By spacetime symmetry,  the $d$-form  flux in a vacuum will be of the form 
\begin{equation}\label{eq:Fd}
\begin{split}
		F_d= \dd A_{d-1}= L^{-1} h\, \mathrm{vol}_{\mathrm{EAdS}_d} = L^{d-1} h \sinh^{d-1} r\, \dd r \wedge \mathrm{vol}_{S^{d-1}}\,;\\
		A_{d-1} = L^{d-1} h\, c(r) \, \mathrm{vol}_{S^{d-1}}  \, ,\qquad c'(r)= \sinh^{d-1}r\,,
\end{split}
\end{equation}
for some constant $h$, which in a quantum gravity theory is expected to satisfy an appropriate quantization condition. Now (\ref{eq:S-brane}) becomes
\begin{equation}\label{eq:hyb}
	S = L^{d-1}\mathrm{vol}_{S^{d-1}} \left[\tau \sinh^{d-1} r -  qh   c(r) \right]\,.
\end{equation}
This is extremized at $r=r_0$ if an only if 
\begin{equation}\label{eq:bounce-hyb}
	\tanh r_0 = \frac{(d-1)\tau}{qh   }\,.
\end{equation}
This condition can be satisfied, and then the tunnel effect is allowed, only if
\begin{equation}\label{eq:QT}
	qh    > (d-1)\tau \,.
\end{equation}
Intuitively, this means that the WZ term, which wants to expand the bubble, wins over the gravitational term, which wants to crush it. 

We can now uplift these simple $d$-dimensional arguments to higher $D$-dimensional vacua of the (possibly warped) product form AdS$_d\times M$, where $M$ is a compact $(D-d)$-dimensional space, having in mind the string/M-theory models  ($D=10,11$). We assume that the $(d-1)$-dimensional bubble corresponds to a microscopic $p$-brane of the form $S^{d-1}\times \Sigma$, where $\Sigma\subset M$ is an internal closed $(p-d+2)$-dimensional submanifold, possibly supporting non-trivial world-volume fields. By invoking again the SO($d$) symmetry, the internal cycle  $\Sigma$ as well as the world-volume fields cannot depend on the position along the external $S^{d-1}$. Hence, they must satisfy some internal equations of motion, decoupled from the external ones, which fix them to some particular configuration. One is then reduced to a $(d-1)$-dimensional action of the form \eqref{eq:S-brane}, with $\tau$ and $q $ fixed by the on-shell internal  $p$-brane configuration. After having computed $\tau$ and $q$ in this way, one can then check whether the instability condition \eqref{eq:QT} is satisfied or not. Of course we expect that \eqref{eq:QT} never holds for supersymmetric vacua. In the following section we will highlight the microscopic geometric origin of this stability in the case of D-brane bubbles in AdS$_4$-vacua. 

We also remark that  \eqref{eq:QT} admits an alternative interpretation. Put the Lorentzian version of the same $(d-2)$-brane along a $\mathbb{R}^{1,2}$ slice of  the Lorentzian  AdS$_d$ in Poincar\'e coordinates. If \eqref{eq:QT} holds, the brane will experience a run-away potential, which will push it to the AdS boundary. This brane ejection effect is indeed often adopted as an alternative criterion to detect the instability of an AdS vacuum; it was argued for holographically in \cite{gaiotto-t,bena-pilch-warner} and by using \eqref{eq:QT} in \cite[Sec.~5.1]{apruzzi-deluca-gnecchi-lomonaco-t}. In the last two references, the role of the dependence of $\tau$, $q$ on the internal coordinates was emphasized.

% subsection inst (end)

\subsection{Brane bubbles in type II \texorpdfstring{$\mathrm{AdS}_4$}{AdS4} vacua} % (fold)
\label{sub:brane}

We now focus on D-brane bubbles in type II  vacua of the form  AdS$_4\times_\mathrm{\s{W}} M_6$, with the metric given by $\d s² = e^{2A}\,\d s²_{\mathrm{AdS}_4} + \d s²_{M_6}$. We will work in string units $ 2\pi\sqrt{\alpha'}=1$.
We begin by recalling that the conditions for such a vacuum to have $\caln=1$ supersymmetry can be written as \cite{gmpt2}
\begin{subequations}\label{eq:psp-eq}
	\begin{align}
		\label{eq:psp-eq1}&\dd_H (\ee^{2A- \phi} \mathrm{Re} \Phi_\mp) =0 \,,\\
		\label{eq:psp-eq2}&\dd_H (\ee^{3A- \phi} \Phi_\pm )= \frac2L \ee^{2A-\phi} \mathrm{Re} \Phi_\mp \,, \hspace{4cm} \overset{\text{\tiny{IIA}}}{\text{\tiny{IIB}}}\\
		\label{eq:psp-eq3}&\dd_H (\ee^{4A-\phi} \mathrm{Im} \Phi_\mp ) =\frac3L \ee^{3A-\phi} \mathrm{Im} \Phi_\pm - \ee^{4A}* \lambda f\,,
	\end{align}
\end{subequations}
where $*$ is the Hodge-star along $M_6$. The $\Phi_\pm$ are polyforms that obey a certain algebraic \textit{purity} condition, defined as bilinears of the internal supersymmetry parameters $\eta^a$. We use the normalization $ \| \eta^a_+ \|^2=8 \ee^A$, and $\Phi_\pm$ such that $(\bar\Phi_\pm,\Phi_\pm)=8\ii$; this is convenient but not necessary.  $f=\sum_k f_k$ is a polyform collecting the internal RR fluxes $f_k$, $H$ is the NSNS flux, $\lambda \alpha_k \equiv (-1)^{\lfloor k/2 \rfloor} \alpha_k$ and $\dd_H \equiv \dd - H \wedge$. The total RR flux is related to the internal one by
\begin{equation}\label{eq:RR10}
	F= f + \mathrm{vol}_4 \wedge \ee^{4A} * \lambda f \,.
\end{equation}
Crucially for us, the equations (\ref{eq:psp-eq}) can be interpreted as closure of the (generalized) calibrations for D-branes extended in AdS$_4$ along time plus one, two, and three space coordinates respectively \cite{martucci-smyth,koerber-martucci-ads}.

Consider now a Euclidean D$p$-brane bubble of the form $S^3\times\Sigma$, where $\Sigma\subset M_6$ is a $(p-2)$-cycle. $\Sigma$ can support a world-volume two-form flux $\calf$, satisfying the Bianchi identity $\d\calf=H|_{\Sigma}$, where $|_{\Sigma}$ denotes the pull-back to $\Sigma$. By reducing the usual D-brane DBI action on $\Sigma$, we easily get 
\begin{equation}\label{Dtau}
   \tau=2\pi\int_{\Sigma}\d^{p-2}\xi\,\ee^{3A}\sqrt{\det(h+\calf)} \,.
\end{equation}
Here $\xi^i$ are world-volume coordinates along $\Sigma$ and $h_{ij}$ is the pull-back of the string frame  metric along $M_6$. The factor of $\ee^{3A}$  comes about because the brane is extended along three directions in EAdS$_4$.

In order to obtain $ q h $,
we must analogously reduce the D-brane CS-term
\beq\label{CSD}
-2\pi\int_{S^3\times \Sigma} C\wedge\ee^{\calf}
\eeq
where $C$ is the polyform of RR potentials, such that $F=\d_H C$.  We can take
\begin{equation}
	C = c_\mathrm{int} + L^4 c(r) \mathrm{vol}_{S^3}\wedge \ee^{4A} * \lambda f
\end{equation}
with $\dd c_\mathrm{int}= f$ (which will not play any role) and $c(r)$ as in (\ref{eq:Fd}). By reducing \eqref{CSD} along $\Sigma$ and matching it against   \eqref{eq:S-brane} with $A_{d-1}$ as in \eqref{eq:Fd}, we get 
\beq\label{probeqf}
qh    =2\pi L\int_{\Sigma}e^{4A}*\lambda f\wedge\ee^{\calf}=6\pi\int_{\Sigma}e^{3A-\phi}\Im\Phi_{\pm}\wedge\ee^\calf\, .
\eeq
In the second step, we used \eqref{eq:psp-eq3}.

We now recall that $\Phi_\pm$ can be regarded as D-brane calibrations \cite{martucci-smyth,koerber-martucci-ads}. This allows us to write the bound:
\beq
\label{eq:Phi-cal}
	\d^{p-2}\xi\,\sqrt{\det(h+\calf)} \ge  \left[\mathrm{Re}(\ee^{\ii \theta}\Phi_\pm)|_\mathrm{\Sigma}\wedge\ee^{\calf}\right]_{\rm top} \quad~~~\text{(for any $\theta$)}
\eeq
where the inequality applies to the coefficients of the local top-form $\d^{p-2}\xi$ on $\Sigma$. By using \eqref{eq:Phi-cal} inside  \eqref{probeqf} and recalling \eqref{Dtau}, we get the bound 
\beq\label{qtauf}
qh\leq 3\tau
\eeq
which shows that \eqref{eq:QT} (with the correct numerical coefficient $d-1=3)$ is indeed never attained in these vacua. 

Note also that the calibration appearing in \eqref{probeqf} corresponds to the phase   $e^{\ii\theta}=-\ii$ in \eqref{eq:Phi-cal}. Precisely with such a choice $\Re(e^{\ii\theta}\Phi_\pm)$ is  the calibration for a domain-wall-like D-brane filling the Poincar\'e $\mathbb{R}^{1,2}$ slice of the Lorentzian   AdS$_4$. Such a configuration is supersymmetric precisely when \eqref{eq:Phi-cal}  is saturated. In this case  $qh= 3\tau$ and then we have threshold stability under the brane ejection mechanism discussed in the previous subsection.

This argument clearly illustrates how  the non-perturbative stability of supersymmetric AdS vacua under nucleation of probe Euclidean  brane bubbles is controlled by calibrations. Hence, it is natural to guess that any more  general stability argument involving backreacting branes, which possibly contribute to both the classical background and  the semiclassical nucleated bubbles, must crucially depend on the existence of appropriate calibration structures. The following sections will show how this guess is precisely realized.

% section bubbles (end)

\section{Stability in four-dimensional (super)gravity} % (fold)
\label{sec:rev}

In order to extend the above stability argument beyond the probe regime, we will adapt methods successfully used in gravity without extra dimensions.

The method stems from the expectation \cite{deser-teitelboim,grisaru-stab} that a supersymmetric state in supergravity saturates a quantum BPS bound of the schematic form
\beq
I\equiv \langle\{Q,Q\}\rangle\geq 0 \,, 
\eeq
where $Q$ is any of the real supercharges preserved by the supersymmetric state. One may then loosely conclude that  the supersymmetric state must be stable, or at most threshold-decay to other supersymmetric states saturating the same bound.

A more precise realization of this formal positivity argument  was given by Witten \cite{witten-pe} and Nester \cite{nester}. The first application in these papers was to the Minkowski vacuum in Einstein gravity, resulting in a new proof of the result in \cite{schoen-yau}. Indeed the argument is sufficiently robust that it may be applied to theories without supersymmetry, in which case the spinors involved in the proof can just be taken to be auxiliary objects. The proof was later extended to non-zero cosmological constant $\Lambda$ in gravity and supergravity \cite{Abbott:1981ff,gibbons-hull-warner,hull-positivity}, and later to various models in different dimensions, which may also arise as EFTs or consistent truncations of string/M-theory compactifications. However, somewhat surprisingly, no successful attempt has so far been made to either apply this approach directly to 10/11-dimensional string/M-theory models, or to include local backreacting brane sources in the argument.
 
In the next sections we will explicitly fill these gaps. Since the discussion will inevitably be quite technical, in this section we will first provide a self-contained review of some key points of this formalism in simpler four-dimensional settings.

%%%%%%%%%%%%%%%%%%%%%%%%%%%%%%%%%%%%%%%%%%%%%%%%%%%%%%%%%%%%%%

\subsection{\texorpdfstring{$\mathrm{Mink}_4$}{Mink4} stability from  supergravity}\label{sub:mink4}

Consider first the minimal four-dimensional supergravity action
\beq
\frac{1}{2}\int\d^4 x\sqrt{-g}\left(R+\ii\bar\psi_\mu\gamma^{\mu\nu\rho}D_\nu\psi_\rho\right)
\eeq
where the gravitino $\psi_\mu$ is in the Majorana representation, and we work in Planck units $8\pi G=M^{-2}_{\rm P}=1$.  
As reviewed in App.~\ref{app:details}, by applying the standard Noether procedure we obtain the following formula 
for the supercharge associated with a given  supersymmetry generator $ε$: \beq\label{4dQ}
Q(ε)=-\ii\int_{\del\Sigma} \bar{ε}\,\gamma_5\gamma_\mu\psi_\nu\,\d x^\mu\wedge \d x^\nu
\eeq
where $\gamma_5\equiv\ii \gamma^{\underline{0123}}$ is the four-dimensional chiral operator and   $\del\Sigma$ is the boundary of a three-dimensional non-timelike hypersurface. Now, the Dirac brackets of two such generators must give
$\{Q(ε'),Q(ε)\}=\delta_{ε'}Q(ε)$, which can be computed from $\delta_ε\psi_μ =D_\muε$. Hence, the usual quantum argument  suggests that the positive definite quantity associated with $Q(ε)$ can be identified with $I(ε)\equiv\delta_{ε}Q(ε)$, with $ε$ now considered  as a commuting spinor. This logic leads to the following  quantity:  \beq\label{4dI}
I(ε)= \int_{\del\Sigma} * E_2\,,
\eeq
where
\beq \label{eq:E2}
E_2\equiv -\frac12\,\bar{ε} \gamma_{\mu\nu}{}^\rho D_\rhoε\,\d x^\mu\wedge \d x^\nu\,.
\eeq
Note that in \eqref{4dQ} and $\delta_ε\psi_μ=D_\muε$ the spinor $ε$ is  anticommuting, while in \eqref{eq:E2} it must be considered as a commuting spinor. 

On asymptotically Minkowski spacetimes, one can choose $\Sigma$ to reduce asymptotically to a  plane with radial  coordinate $r$, and identify $\del\Sigma$ with a sphere  of constant radius $R\rightarrow\infty$. If one imposes that $ε$  asymptotically reduces to a constant spinor $ε_0$ up to $\calo(1/r)$ terms, then $I(ε)$ coincides with the Nester--Witten energy \cite{nester}, which is still valid in the presence of additional matter coupled to the minimal gravity multiplet through an energy-momentum tensor $T^{\text{(mat)}}_{\mu\nu}$. More precisely, in an appropriate normalization, we can make the identification 
 \beq\label{IPrel}
 I(ε)=- k_0^\mu P_\mu\,,
 \eeq 
where $P_\mu$ is the four-momentum of the system and
\beq\label{eq:k0-eps}
 k^\mu\equiv \bar{ε}\gamma^\muε\quad\Rightarrow\quad   k_0^\mu\equiv \bar{ε}_0\gamma^\muε_0\,. 
\eeq
In particular, $I(ε)=I(ε_0)$: it only depends on the asymptotic value of the spinor.
Note that here we are using a Majorana  spinor $ε$ (as for instance in \cite{boucher}), while in \cite{nester,witten-pe} a  complex $ε$ is used. This implies that in our case $k^\mu$ is a future-pointing null vector. However, since $ε_0$ is arbitrary,   $k_0^\mu$ spans the entire future light-cone of the asymptotic flat space. 
This means that, if $ I(ε)\geq 0$ for any $ε_0$, then $P^\mu$ must be necessarily non-spacelike and future-pointing. In particular, the energy is non-negative in any frame.

In order to prove that $I(ε)\geq 0$ if $T^{\text{(mat)}}_{\mu\nu}$ satisfies the dominant energy condition, we now  assume that $\Sigma$ is a regular space-like surface as in  \cite{witten-pe,nester}. (Horizons may be included as in \cite{Gibbons:1982jg} and the discussion may be generalized to null surfaces $\Sigma$ following \cite{nester_israel}.) Then by using Stokes' theorem and the identity $\bar{ε} γ_{μνρ} D^ν D^ρ ε=G_{μν} k^\nu$ (which we will show in detail in sec.~\ref{sub:Mmain}), we can write \eqref{4dI} as follows:
\begin{equation}\label{eq:int-af}
\begin{aligned}
	I(ε) &= \int_Σ \d * E_{(2)}  
	= \int_Σ ∇^ν E_{μν} * \d x^μ=  \int_Σ \left( D^ν \bar{ε}γ_{μνρ}D^ρ ε +\frac12G_{μν} k^\nu \right)* \d x^μ\\
	&\overset{\textrm{on-shell}}{=}   \int_Σ D^ν \bar{ε}γ_{μνρ}D^ρ ε n^μ \textrm{vol}_Σ +  \frac12\int_Σ T_{μν}^\textrm{(mat)} k^ν n^μ \textrm{vol}_Σ\,.
\end{aligned}
\end{equation} 
In the last line we have rewritten $* \d x^ν|_\Sigma = \textrm{vol}_Σ n^ν$, where $n^ν$ is a time-like unit vector orthogonal to the surface $Σ$ and $\textrm{vol}_Σ$ is the induced volume form on $\Sigma$, and we have imposed Einstein's equations $G_{μν} =  T_{μν}^\textrm{(mat)}$.
The last term of \eqref{eq:int-af} is manifestly non-negative if $T_{μν}^\textrm{(mat)}k^ν n^μ\geq 0$ for any null future-directed vector $k^\mu$ and then, by linearity, for any non-space-like vector $k^\mu$. This is precisely the definition of the dominant energy condition, which we are assuming.  In order to prove that $I(ε)\geq 0$, it then remains to show that also the first term in the last line of \eqref{eq:int-af} is non-negative. 

At this point it is convenient to  pick an adapted vielbein  $e^\s{A}=e^\s{A}_\mu\d x^\mu=(e^{\underline{0}},e^a)$ (with spatial flat indices $a\in \{\underline{1},\underline{2},\underline{3}\}$) and the dual frame $e_\s{A}=e_\s{A}^\mu\del_\mu=(e_{\underline{0}},e_a)$  such that $e^{\underline{0}}|_\Sigma=0$ and then $n^\mu\equiv e_{\underline{0}}^\mu$.  The first term in the last line of \eqref{eq:int-af}  can then be decomposed into:
\begin{equation}\label{eq:witten-4}
	\int_Σ D^a \bar{ε}\gamma_{\underline{0}ab}D^b ε \textrm{vol}_Σ = \int_Σ \left(D^a ε\right)^\dagger \left(D_a ε\right) \textrm{vol}_Σ - \int_Σ |\gamma^a D_a ε|^2\textrm{vol}_Σ .
\end{equation}
The first contribution is manifestly positive; the second vanishes upon imposing the \emph{Witten condition}:
\begin{equation}\label{4dWcond}
	\gamma^a D_a ε = 0.
\end{equation}
This has a natural interpretation from the supergravity viewpoint \cite{hull-positivity}: it can be regarded as following from a ``transverse'' gauge choice  $\gamma^a\psi_a=0$ \cite{Das:1976ct,Deser:1977ur} on the gravitino. In other words, the last term in \eqref{eq:witten-4} can be  associated with ``longitudinal'' unphysical degrees of freedom, which can be gauged away. 

Let us for the moment assume  that \eqref{4dWcond}  can be solved  for any choice of the spatial surface $\Sigma$ and asymptotic $ε_0$. Combining \eqref{eq:int-af} and \eqref{eq:witten-4}, we can then conclude that $I(ε)\geq 0$. Moreover, if $I(ε)=0$,  $D_a ε$ needs to vanish; if this is true for any $ε_0$, by varying $\Sigma$ we find four constant spinors on our spacetime, which implies that it is  Minkowski. Since the only spacetime where $I(ε)=0$ for any $ε_0$ is Minkowski, and $I(ε)$ is conserved (it does not depend on $\Sigma$), Minkowski space cannot evolve into anything else: it is stable.

The remaining key step is then to prove that \eqref{4dWcond} always admits a solution for the given boundary condition $ε→ε_0 + \mathcal{O}\left(1/r\right)$. This question can be addressed at various levels of mathematical rigor \cite{witten-pe,Parker:1981uy,Gibbons:1982jg}, but for our purposes we will just focus on the most crucial requirement: the elliptic operator $\gamma^a D_a$ must have no normalizable zero modes. If this holds,  $\gamma^a D_a$  can be inverted into a Green's operator. We can then start with any trial $ε_1$ asymptotic to $ ε_0$ and find a normalizable  solution $ε_2$ to $\gamma^a D_a ε_2 =-\gamma^a D_a ε_1$; now $ε_1+ ε_2$ solves \eqref{4dWcond} \cite{witten-pe,Parker:1981uy}.
The absence of a normalizable zero mode $ε_{\rm zm}$ of $\gamma^a D_a$  can be understood as follows. Such an $ε_{\rm zm}$ would decrease sufficiently fast to make $I(ε_{\rm zm})$ as defined in \eqref{4dI} vanish. But then, by rewriting
$I(ε_{\rm zm})$ as in \eqref{eq:int-af} and using  $\gamma^a D_aε_{\rm zm}=0$ inside the identity \eqref{eq:witten-4}, we would get 
\beq\label{zmcond}
\int_Σ \left(D^a ε_{\rm zm}\right)^\dagger \left(D_a ε_{\rm zm}\right)+\frac12\int_Σ T_{μν}^\textrm{(mat)} k_{\rm zm}^ν n^μ \textrm{vol}_Σ=0\,.
\eeq
 This  identity is clearly impossible  unless $D_a ε_{\rm zm}=0$ for any $\Sigma$, and then $ε_{\rm zm}\equiv 0$ (since  $ε_{\rm zm}$ must vanish asymptotically). We will also apply the same kind of argument to more complicated settings.

\subsection{\texorpdfstring{$\mathrm{AdS}_4$}{AdS4} stability}
\label{sub:AdS4}

The logic outlined  in the previous subsection can be adapted to AdS vacua in theories of gravity and supergravity \cite{Abbott:1981ff,gibbons-hull-warner,hull-positivity}. The simplest, ``minimal'' model is obtained by adding a constant superpotential $W_0$, which generates a negative cosmological constant  $\Lambda=-3|W_0|^2$ (in Planck units). Without loss of generality, we assume that $W_0$  is real and positive, so that $W_0=\sqrt{-\Lambda/3}\equiv 1/L$, where $L$ is the AdS radius. In global coordinates, the AdS metric reads
\beq\label{GAdS}
\d s^2_{\text{AdS}_4}=-\left(1+\frac{\rho^2}{L^2}\right)\d t^2+\left(1+\frac{\rho^2}{L^2}\right)^{-1}\d \rho^2+\rho^2\d\Omega^2\,;
\eeq
at $\rho\to \infty$, the rescaled $\rho^{-2} \d s^2_{\text{AdS}_4}$ induces on the boundary the metric of $\mathbb{R}\times S^2$. 
The formula \eqref{4dQ} for the supercharge is unchanged while the gravitino transformation is modified into $\delta_ε\psi_\mu={\cal D}_\muε$, where
\beq \label{eq:DW-4d}
{\cal D}_\mu\equiv D_\mu+\frac12 W_0\gamma_\mu\,.
\eeq
Recall that an $\varepsilon$ annihilated by ${\cal D}_\mu$ is said to be a \textit{Killing spinor}. AdS has the maximal number of such spinors: they behave asymptotically as \cite[(3.17)]{breitenlohner-freedman} $ε \sim \rho^{1/2} ε_0$, with $ε_0$ projecting on the boundary to a conformal Killing spinor of $\mathbb{R}\times S^2$.
 
Changing $D \to {\cal D}$ everywhere in the argument \eqref{eq:int-af}--\eqref{4dWcond} in the previous subsection now again establishes in this minimal model that $I(\varepsilon)\ge 0$, and that $I(\varepsilon)=0$ only in the AdS vacuum.  The same procedure can be repeated for non-minimal gauged supergravities \cite{gibbons-hull-warner}: the only difference is the appearance of additional contributions to $I(ε)$, also manifestly non-negative, containing the supersymmetry variations of additional fermions in the supersymmetric multiplets. 
(See also  \cite{Hristov:2011ye,Hristov:2011qr} for applications to  the calculation of BPS bounds for non-vacuum states in $\caln=2$ gauged  supergravities.)
However, the different global structure of AdS changes the rest of the argument. The presence of the $\mathbb{R}\times S^2$ timelike boundary makes the spacetime not \textit{globally hyperbolic} --- one cannot predict the future from the data on a Cauchy slice alone, but rather one also needs to know what happens at spatial infinity. A priori this might make it unclear whether $I(\varepsilon)$ is conserved. A natural boundary condition for the metric is to impose that $\rho^{-2}\d s^2$ be conformal to $\mathbb{R}\times S^2$ at the boundary, as happens for the AdS vacuum. In \cite{Hollands:2005wt} it was shown (in a pure gravity model) that with this boundary condition, $I(\varepsilon)$ is equal to the conserved charge ${\cal E}(k_0)$ associated by the covariant phase space formalism \cite{iyer-wald} to the asymptotic isometry $k_0^\mu= \bar \varepsilon_0 \gamma^\mu \varepsilon_0$. The latter are conformal Killing vectors of $\mathbb{R}\times S^2$, and together they generate its conformal isometry group ${\rm SO}(2,3)$. Moreover, ${\cal E}(k_0)$ does not depend on the space slice $\Sigma$ (\ie it is conserved).\footnote{A detailed comparison of $I(\varepsilon)$ to alternative notions of energy in asymptotically AdS spacetimes is also given in \cite{Hollands:2005wt}; see also \cite{Marolf:2012vvz} for a review.} 

That the asymptotically supersymmetric $\varepsilon$ can satisfy the Witten condition (now $\gamma^a {\cal D}_a \varepsilon=0$) was shown in a slightly more general context in \cite[Sec.~V]{amsel-hertog-hollands-marolf}. So we have in fact several non-negative conserved quantities, to which we can apply the argument in the previous subsection. 

The covariant phase space method can also be applied to supersymmetry itself, in which case it reproduces once again the expression \eqref{4dQ} for the supercharge $Q(\varepsilon)$ \cite{Hollands:2006zu}. The formalism is built so that the conserved charge is a Hamiltonian generator for the associated symmetry; from the (super-)Jacobi identity it follows that  $Q(\varepsilon)$ and the  ${\rm SO}(2,3)$ generators $J_{\s{AB}}$ together  form a superalgebra, as expected. In particular \cite{Abbott:1981ff,gibbons-hull-warner,hull-positivity} 
\beq\label{AdS4I}
I(ε)=\frac12J_{\s{AB}}\,\bar{ε}_0\sigma^{\s{AB}}ε_0\, ,
\eeq   
where $\sigma^{ab}\equiv \frac12\gamma^{ab}$ and $\sigma^{a4}=\frac12\gamma^a$. The AdS energy $E_{\rm AdS}\equiv J_{04}$ is obtained by tracing \eqref{AdS4I} over a basis of $ε_0$; it vanishes only when both $T_{μν}^\textrm{(mat)}\equiv 0$ and $D_μ ε \equiv 0$, \ie maximally supersymmetric AdS is the unique $E_{\rm AdS}=0$ configuration of these models.

\subsection{Breaking supersymmetry}
\label{sub:susy-br-4}

As we mentioned at the beginning, the arguments in this section are robust enough that they don't need supersymmetry: the spinors $\varepsilon$ can be auxiliary variables, unrelated to any fermionic symmetry. Indeed, even the first application \cite{witten-pe,nester} was to pure Einstein gravity. One can follow this idea to give stability arguments both for vacua in non-supersymmetric theories, and for supersymmetry-breaking vacua in supersymmetric ones.

A systematic analysis was initiated in \cite{boucher} (and generalized to $d>4$ in \cite{townsend-stab}). To illustrate the idea, consider a model with a single scalar $\phi$ and action 
\beq
\frac{1}{2}\int\d^4 x\sqrt{-g}\left(R- \partial_\mu\phi \partial^\mu \phi - 2 V(\phi)\right)\,.
\eeq
There is no supersymmetry, but we may nevertheless consider spinors $\varepsilon$ that satisfy ${\cal D}'_\mu \varepsilon=0$, with
\begin{equation}
    {\cal D}'_\mu \equiv D_\mu + \frac12 W(\phi) \gamma_\mu \,,
\end{equation}
generalizing \eqref{eq:DW-4d}. Trying to prove an analogue of \eqref{eq:int-af}, one finds that it works if 
\begin{equation}\label{eq:fake}
    V= 2 (\partial_\phi W)^2 - 3 W^2\,.
\end{equation}
This has the same structure of an ${\cal N}=1$ model; but one may also look for such a $W$ in absence of supersymmetry, in which case it is known as a \textit{fake} superpotential. 

The discussion of boundary conditions for $\phi$ is quite intricate; see for example \cite{amsel-hertog-hollands-marolf}. As is well known, near the boundary it behaves like $\alpha_- \rho^{-\Delta_-}+ \alpha_+ \rho^{-\Delta_+}$, where $\Delta_- \le \Delta_+$ are the two roots of $\Delta(\Delta-d+1)= m^2$. For ``fast'' boundary conditions $\alpha_-=0$, the BPS energy again is equal to the conserved energy from the covariant phase space formalism, but for choices of the type $\alpha_+= f(\alpha_-)$ the two differ. In fact there are in general two local solutions $W_-\le W_+$ to \eqref{eq:fake} for each given $V$ around the vacuum; $W_+$ always exists globally, but when $\alpha_-\neq0$ the BPS energy associated to it is infinite. So one has to use $W_-$, but its global existence is guaranteed only for some potentials.

\subsection{Some comments}

We close this section with some remarks. 

\begin{itemize}

\item
Even though the identification of the positive definite $I(ε)$ exploits the supersymmetric structure, the stability argument regards only  purely bosonic configurations. This is not really an issue  at the classical level, since  non-vanishing fermionic profiles do not admit a classical interpretation. Hence, the above stability arguments are complete  as long as we are interested in classical instabilities or semiclassical ones mediated by Coleman--de Luccia instantonic bubbles \cite{coleman-deluccia,Abbott:1985kr}. Indeed, the portion of spacetime created after nucleation of a bubble can be considered as a localized classical excitation of the Mink$_4$ or AdS$_4$ vacuum, preserving the appropriate boundary conditions.

\item These four-dimensional arguments can be applied to string/M-theory models as long as the four-dimensional theory is a consistent EFT and the possible decay processes are describable within the EFT regime of validity. But many examples of  string/M-theory  compactifications to AdS$_4$ do not admit such an EFT description, since for instance the AdS scale $\sqrt{-\Lambda}$ and the KK scale are of the same order.  In such cases, there often exist  some consistent truncations which lead to extended gauged supergravities of the kind considered in  \cite{gibbons-hull-warner}. However, any conclusion on the stability of the vacuum based on these consistent truncations is inconclusive.

\item To our knowledge, the models considered so far in the literature do not include possible charged membranes, which for instance catalyse the kinds of decays discussed   in section \ref{sec:bubbles} (in the probe approximation). Such membranes could be incorporated into a four-dimensional $\caln=1$ EFT as in \cite{Bandos:2018gjp,Lanza:2019xxg} and should be taken into account. Moreover, in the absence of scale separation, such membranes should actually be regarded as higher dimensional branes as in section \ref{sub:brane}.  

\end{itemize}

In the next section we will address some of the open issues raised in the second and third items by explicitly showing how the above stability arguments can be uplifted to string/M-theory. Moreover, we will see how various (backreacting) branes can be taken into account, hence providing a first important set of examples on how such extended charged objects can be consistently incorporated into these kinds of  arguments. 

%%%%%%%%%%%%%%%%%%%%%%%%%%%%%%%%%%%%%%%%%%%%%%%%%%%%%%%%%%%%%%

\section{Positivity and stability in M-theory} % (fold)
\label{sec:m}

We now move on to the (expected)  stability of supersymmetric string/M-theory supergravity backgrounds, including possible localized sources. In this section we will consider the M-theory case, while in the next we will discuss type II theories. The extension of our results to other string theories should be straightforward.      

Some useful properties of eleven-dimensional supergravity are reviewed in App.~\ref{app:details}, in which we also fix our conventions. Here we just recall that the bulk fields are the metric $g_{\s{MN}}$, a three-form potential $C$ with closed four-form field-strength $G=\dd C$,  and a Majorana gravitino $\psi_{\s{M}}$. Furthermore, we will denote the eleven-dimensional gamma matrices by $\Gamma_{\s{M}}$, and we will work in units $\lp=1$, in which the supergravity Einstein--Hilbert term is $2\pi\int R*1$ and the M2-brane tension is $2\pi$.

Let us follow the same steps of section \ref{sec:rev}. First,  one can again apply the Noether procedure reviewed in App.~\ref{app:noether}, to get the following expression for the supercharge: 
\begin{equation}\label{Qdef}
	Q(\epsilon)=-\int_{\del S}\bar\epsilon\, \Gamma_{(8)}\wedge \psi\,,
\end{equation}
where (commuting)  Majorana spinor $\epsilon$ defines an arbitrary supersymmetry transformation and $\del S$ is the boundary of a ten-dimensional  hypersurface $S$. We are using the intrinsic notation where $\psi\equiv \psi_{\s{M}}\dd x^{\s{M}}$ and
\begin{equation}\label{eq:G(p)}
	\Gamma_{(p)}\equiv \frac1{p!} \Gamma_{\s{M_1\cdots M_p}}\dd x^{\s{M_1}} \wedge \cdots \wedge \dd x^{\s{M_p}}\,.
\end{equation}
The supersymmetry transformation of the gravitino reads 
\begin{equation}
	\delta \psi_{\s{M}} = {\mathcal D}_{\s{M}} \epsilon+\calo(\psi^2)\,,
\end{equation}
where $\epsilon$ is a Majorana spinor and
\begin{equation}\label{McalDdef}
{\mathcal D}_{\s{M}}\equiv D_{\s{M}}+\frac1{12}(G\Gamma_{\s{M}}-G_{\s{M}})
= D_{\s{M}}+\frac1{24}(- \Gamma_{\s{M}} G + 3 G\Gamma_{\s{M}})\,,
\end{equation}
with $D_{\s{M}}$ being the usual spinorial covariant derivative. In this expression, $G$ should actually be written as $G_{\slash}\equiv \frac1{4!} G_{\s{MNPQ}} \Gamma^{\s{MNPQ}}$ (often also denoted by $\slashed{G}$), the bispinor associated to the four-form $G$ under the Clifford map ${}_{\slash}$ in  \eqref{eq:clifford}. 
To get cleaner spinorial equations, we will often omit the slash symbols, hoping it will be clear from context whether we are referring to a form or to the associated bispinor. For instance, in \eqref{McalDdef} we also have $G_{\s{M}} \equiv (\iota_{\s{M}} G)_{\slash} = \frac1{3!} G_{\s{MNPQ}} \Gamma^{\s{NPQ}}$ -- again see App.~\ref{app:details} for more details on our notation. 

The argument outlined in section \ref{sub:mink4} leads to the following natural candidate for a positive definite quantity associated with $\epsilon$:
\begin{equation}\label{genEnergy}
I(\epsilon)\equiv \int_{\del S} *E_2
\end{equation}
where 
\begin{equation}\label{eq:ME2}
E_2\equiv  -\frac12\,\bar\epsilon\,\Gamma_{\s{MN}}{}^{\s{P}}{\mathcal D}_{\s{P}}\epsilon\,\dd x^{\s{M}}\wedge \dd x^{\s{N}}\,.
\end{equation}
This two-form is the  eleven dimensional counterpart of  \eqref{eq:E2}.
%\eqref{eq:int-af} or  \eqref{eq:E-AdS} 

These formulas have already been exploited together with the usual quantum supersymmetry argument to formally argue for some BPS bounds as in \cite{hull-charges} -- see also \cite{legramandi-martucci-t} for a more recent discussion in terms of calibrations which is more directly  relevant for what follows. For this reason, we will refer to \eqref{genEnergy} as the {\em BPS energy}, having in mind a non-negative linear combination of energy and possible central charges that vanishes when supersymmetry is preserved. However, even though it is expected, positivity of the BPS energy \eqref{genEnergy} has not been proved to date. This is precisely one of our main goals.

In order to prove positivity, as in the four-dimensional case we assume that $S$ is a regular space-like hypersurface and use Stokes' theorem
\begin{equation}\label{genEnergy2}
I(\epsilon) =\int_{S} \d *E_2=\int_S\, \nabla^{\s{M}}E_{\s{NM}}\,*\dd x^{\s{N}}\, .
\end{equation}
Horizons and asymptotically null hypersurfaces $S$ may be also considered following \cite{gibbons-hawking-horowitz-perry} and  \cite{nester_israel} respectively, but we will not do it here. On the other hand, we will   allow for  M2 and M5 branes, showing how  to properly take into account their backreaction.

\subsection{The main identity}
\label{sub:Mmain}

Our next crucial step is to compute the divergence appearing in (\ref{genEnergy2}). We start by presenting the final result: 
\begin{equation}\label{eq:nablaE}
	 \framebox
	 {$\nabla_\s{M} E^{\s{MN}} = \overline{{\mathcal D}_\s{M} \epsilon} \Gamma^{\s{MPN}} {\mathcal D}_\s{P} \epsilon +\frac12\cale^{\s{NP}}K_\s{P}+\frac14 \bar \epsilon[\dd x^\s{N} \wedge (-\d G +\d* G+\frac12G\wedge G)]_{\slash}\epsilon\,,$}
\end{equation}
where we have introduced the future-pointing causal vector
\beq\label{MKdef}
K\equiv K^\s{M}\del_\s{M}=\bar\epsilon\,\Gamma^\s{M}\epsilon\,\del_\s{M}\,,
\eeq
and $\cE_{\s{MN}}$ denotes the bulk contribution to the Einstein equations:
\beq\label{eq:geom}
\begin{aligned}
\cE_{\s{MN}}&\equiv R_{\s{MN}}-\frac12 g_{\s{MN}} R -\frac12 T_{\s{MN}}^{\s{(G)}} \,\\
&\quad~~~\text{where}\quad~~~~ T_{\s{MN}}^{\s{(G)}}\equiv G_\s{M} \cdot G_\s{N} -\frac12 |G|^2 g_{\s{MN}}\,.
\end{aligned}
\eeq
Note that \eqref{eq:nablaE} holds identically for {\em any off-shell} configuration of $g_\s{MN}$ and $G$, in the sense that one does not need to impose either the equations of motion or the Bianchi identities. 

While until here our logic followed quite closely the steps in section \ref{sec:rev}, the proof of (\ref{eq:nablaE}) is quite specific to M-theory. We present here the main steps of the argument, leaving several details to App.~\ref{app:m}. We begin with 
\begin{equation}\label{eq:nablaE-0}
\begin{split}
	\nabla_\s{M} E^{\s{MN}} =& -\nabla_\s{M} (\bar \epsilon \Gamma^{\s{MNP}} {\mathcal D}_\s{P} \epsilon)= -\overline{D_\s{M} \epsilon} \Gamma^{\s{MNP}} {\mathcal D}_\s{P} \epsilon 
	+ \bar \epsilon\, \Gamma^{\s{NMP}} D_\s{M} {\mathcal D}_\s{P} \epsilon \\
	=& -\overline{{\mathcal D}_\s{M} \epsilon} \Gamma^{\s{MNP}} {\mathcal D}_\s{P} \epsilon 
	-\frac1{24} \bar \epsilon\, A^{\s{NP}} D_\s{P} \epsilon \\
	&- \bar \epsilon\, \left[\Gamma^{\s{MNP}} \left(D_\s{M} D_\s{P} 
	+\frac1{24} [D_\s{M}, -\Gamma_\s{P}  G+ 3 G \Gamma_\s{P}]\right)+ \frac1{24^2}Q^\s{N} \right]\epsilon\,,
\end{split}
\end{equation}
where
\begin{equation}\label{eq:AB-m}
	\begin{split}
	A^{\s{NP}} &= (-G \Gamma_\s{M} + 3 \Gamma_\s{M} G) \Gamma^{\s{MNP}}- \Gamma^{\s{NPM}}(-\Gamma_\s{M} G + 3 G \Gamma_\s{M}) \, ,\\ 
	Q^\s{N} &=(-G \Gamma_\s{M} + 3 \Gamma_\s{M} G) \Gamma^{\s{MNP}}(- \Gamma_\s{P} G + 3 G \Gamma_\s{P})\,.
	\end{split}	
\end{equation}
We show in App.~\ref{app:m} that $A^{\s{NP}}=0$. The first term on the last line of (\ref{eq:nablaE-0}) can be evaluated recalling that the commutator of two spinorial covariant derivatives involves the Riemann tensor, and some relatively standard gamma matrix identities:
\begin{align}%\label{eq:GMN}
	\nonumber
	\Gamma^{\s{MNP}}[D_\s{N},D_\s{P}] &= \frac14 R^{\s{AB}}{}_{\s{NP}} \Gamma^{\s{MNP}} \Gamma_{\s{AB}}
	= \frac14  R^{\s{AB}}{}_{\s{NP}} (\Gamma^{\s{MNP}}_{\s{AB}} + 6 \delta^{[\s{M}}_\s{A} \Gamma^{\s{NP}]}_\s{B}-6 \delta^{[\s{M}}_\s{A} \delta^{\raisemath{1.5pt}{\s{N}}}_\s{B}\Gamma^{\raisemath{1.5pt}{\s{P}]}}) \\
	\label{eq:GMN}
	&= \left(R^{\s{MN}}-\frac12 R g^{\s{MN}}\right) \Gamma_\s{N}\,.
\end{align}
The commutator term on the last line in (\ref{eq:nablaE-0}) is computed in the appendix as
\begin{equation}
	\Gamma^{\s{NMP}}[D_\s{M}, -\Gamma_\s{P}  G+ 3 G \Gamma_\s{P}] = 6 ( - \dd x^\s{N} \wedge \dd G + \iota^\s{N} * \dd *  G)_{\slash}\,.
\end{equation}
Finally we also show in  App.~\ref{app:m}  that $Q^\s{N}$ evaluates precisely to the quadratic terms in the equations of motion of the metric and $G$, leading to (\ref{eq:nablaE}).

\subsection{Positivity in the absence of branes}
\label{sub:Mpos}

We are now in a position to present our positivity argument, following the four-dimensional case reviewed in section \ref{sec:rev}. In this subsection we make the simplifying assumption that no M2 or M5 branes are present or can be nucleated. The inclusion of  branes will be discussed in the following subsection.

In the absence of branes, the equations of motion and Bianchi identities of the supergravity fields are 
\begin{equation} \label{eq:Geom}
	\cale_{\s{MN}}=0 \, ,\qquad \d G =0 \, ,\qquad \d*G+\frac12 G\wedge G=0\,.
\end{equation}
Hence, the identity \eqref{eq:nablaE} reduces on-shell  to 
\beq
\nabla_\s{M} E^{\s{MN}}\  \overset{\textrm{on-shell}}{=} \  \overline{{\mathcal D}_\s{M} \epsilon} \Gamma^{\s{MPN}} {\mathcal D}_\s{P} \epsilon\quad~~~\text{(no branes)}\,, 
\eeq
and the BPS energy \eqref{genEnergy2} becomes
\beq\label{IMonshell}
I_{0}(\epsilon)= \int_S\, \overline{{\mathcal D}_\s{M} \epsilon} \Gamma^{\s{MN}}{}_\s{P} {\mathcal D}_\s{N} \epsilon\, n^\s{P}\text{vol}_S\,,
\eeq
where $\text{vol}_S$ is the volume form induced on $S$ and $n=n^\s{M}\del_\s{M}$ is a future-pointing unit vector orthogonal to $S$. We can now proceed as in four dimensions, by using an adapted vielbein 
\beq\label{Made}
e^\s{A}=(e^{\underline{0}},e^a)\quad\text{with}\quad  e^{\underline{0}}|_S=0\,,
\eeq 
and the dual frame $e_\s{A}=(e_{\underline{0}},e_a)$ with $e_{\underline{0}}=n$. 
Now \eqref{IMonshell} can be rewritten as 
\begin{equation}\label{eq:E-square}
	\begin{aligned}
	I_{0}(\epsilon)&= \int_S \text{vol}_S\, \left[({\mathcal D}^a\epsilon)^\dagger({\mathcal D}_a\epsilon)-|\Gamma^a {\mathcal D}_a\epsilon|^2\right] \,.
	\end{aligned}	
\end{equation}
By choosing an $\epsilon$ that satisfies the M-theory version of the Witten condition
\begin{equation}\label{eq:w}
	\Gamma^{a}{\mathcal D}_a\epsilon=0\,,
\end{equation}
we see that the BPS energy is manifestly non-negative: $I_{0}(\epsilon)\geq 0$. Furthermore,  $I_{0}(\epsilon)= 0$ if and only if $\cald_a\epsilon=0$. Since $S$  can be freely deformed, this condition actually requires that  $\cald_\s{M}\epsilon=0$. Hence, the BPS energy is vanishing if and only if $\epsilon$ satisfies the supersymmetry equations. 

Note that, as in the four-dimensional case, we can again associate \eqref{eq:w} with a transverse gauge-fixing condition  $\Gamma^a\psi_a=0$ for the gravitino. Furthermore, one can adapt the argument around \eqref{zmcond} to argue that  for any $S$ the operator $\Gamma^a\cald_a$ does not admit any normalizable  zero  mode $\epsilon_{\rm zm}$. Indeed, by combining \eqref{eq:E-square} with $I_0(\epsilon_{\rm zm})=0$ and $\Gamma^a\cald_a \epsilon_{\rm zm}=0$ one gets $\cald_a\epsilon_{\rm zm}= 0$. Since $\epsilon_{\rm zm}$ must vanish as one approaches the boundary $\del S$, we then expect that the conditions $\cald_a\epsilon_{\rm zm}= 0$ impose that $\epsilon_{\rm zm}\equiv 0$, at least for reasonable asymptotic structures. This supports the expectation that \eqref{eq:w} indeed admits a solution, although an explicit proof of this result would  require a more detailed case by case investigation which is beyond the scope of the present paper.

\subsection{Inclusion of M2 branes}
\label{sec:M2}

We would now like to allow M2 or M5 branes to be present in the background configuration, or to be generated by possible fluctuations around it. 

We first assume the presence of only M2-branes. This means that the Bianchi identity $\d G=0$ is still satisfied, while the bulk equations of motion are modified.  The bosonic action of an M2-brane  is
\beq\label{M2action}
S_{\text{\tiny(M2)}}=-2\pi\int_\cC\d^3\sigma\sqrt{-h}+2\pi\int_\cC C\,,
\eeq
where $h_{\alpha\beta}\equiv g_{\s{MN}}\del_\alpha X^\s{M}\del_\beta X^\s{N}$ is the metric induced on the world-volume $\calc$  by the embedding $\sigma^\alpha\mapsto X^\s{M}(\sigma)$.
The WZ term in \eqref{M2action} provides a localized source term to the $C$ equations of motion:
\beq\label{GeomM2}
\d *G+\frac12G\wedge G=\delta^{(8)}(\cC)\,,
\eeq
where $\delta^{(8)}(\cC)$  is a delta-like eight-form such that $\int_\calc C\equiv\int C\wedge \delta^{(8)}(\cC)$ for any three-form $C$.\footnote{More generically, for any $q$-dimensional submanifold $\Sigma$ of a $d$-dimensional space, $\delta^{(d-q)}(\Sigma)$ is defined by $\int_\Sigma \omega=\int \omega\wedge \delta^{(d-q)}(\Sigma)$ for any $q$-form $\omega$.} Analogously, the Nambu--Goto term of \eqref{M2action}  enters the Einstein equations by a localized source term $T^{\s{MN}}_{\text{\tiny (M2)}}$, such that:
\beq\label{M2Einst}
\cale^{\s{MN}}=\frac12 T^{\s{MN}}_{\text{\tiny (M2)}}\,,
\eeq
where $\cale^{\s{MN}}$ is defined in \eqref{eq:geom}. 
In order to describe $T^{\s{MN}}_{\text{\tiny (M2)}}$, it is convenient to pick an adapted  vielbein
\beq\label{M2vielbein}
e^\s{A}=(e^{\underline \alpha},e^{\tilde a})\quad~~~\text{with  $ e^{\tilde a}|_\calc=0$}\,,
\eeq
and the corresponding dual frame $e_\s{A}=e_{\s{A}}^\s{M}\del_\s{M}=(e_{\underline \alpha},e_{\tilde a})$, with $A=0,\ldots, 10$, $\alpha=0,1,2$ and $\tilde a=3,\ldots, 8$. Then one can verify that
\beq\label{M2EM}
T^{\s{MN}}_{\text{\tiny (M2)}}=\eta^{\underline{\alpha\beta}}\,e_{\underline\alpha}^\s{M} e_{\underline\beta}^\s{N}*[e^{\underline{012}}\wedge \delta^{(8)}(\cC)]\,,
\eeq
see App.~\ref{app:M2}.
Note that  the right-hand side of \eqref{GeomM2} and \eqref{M2Einst}  can be immediately generalized to  multiple M2-branes  by summing over the corresponding contributions. Hence, we can focus on a single M2-brane without loss of generality. 

By using \eqref{GeomM2} and \eqref{M2Einst} (together with $\d G=0$) in the identity \eqref{eq:nablaE}, we now get 
\beq\label{DEM2}
\begin{aligned}
\nabla_\s{M} E^{\s{MN}} &\overset{\textrm{on-shell}}{=}  \overline{{\mathcal D}_\s{M} \epsilon} \Gamma^{\s{MPN}} {\mathcal D}_\s{P}\epsilon+\frac14 T^{\s{NP}}_{\text{\tiny (M2)}}K_\s{P}+\frac14\bar\epsilon[\d x^\s{N}\wedge \delta^{(8)}(\calc)] \epsilon\\
&\quad =\quad\overline{{\mathcal D}_\s{M} \epsilon} \Gamma^{\s{MPN}} {\mathcal D}_\s{P}\epsilon+\frac14 T^{\s{NP}}_{\text{\tiny (M2)}}K_\s{P}+\frac14 \big(*\big[\d x^\s{N}\wedge \delta^{(8)}(\cC)\big]\big)\cdot \Omega^{\text{\tiny (M2)}}\,, \end{aligned}
\eeq
where we have introduced the  two-form
\beq\label{OmegaM2}
\Omega^{\text{\tiny (M2)}}\equiv \bar\epsilon\,\Gamma_{(2)}\epsilon=\frac12\bar\epsilon\,\Gamma_{\s{MN}}\epsilon\,\d x^\s{M}\wedge \d x^\s{N}\,.
\eeq
By using \eqref{DEM2} inside 
\eqref{genEnergy2}  we find that the total BPS energy can be split into bulk and brane contributions:  
\beq\label{IsplitM2}
I(\epsilon)=I_0(\epsilon)+I_{\text{\tiny (M2)}}(\epsilon)\,.
\eeq
Here $I_0(\epsilon)$ is defined  in \eqref{IMonshell} and represents the bulk contribution. On the other hand, the brane contribution 
\beq\label{M2BPSenergy}
I_{\text{\tiny(M2)}}(\epsilon)= \frac14\int_{\cC\cap S}\left(K^{\underline{0}}\,{\rm vol}_{\cC\cap\Sigma}-\Omega^{\text{\tiny(M2)}}\right)
\eeq
 comes  from the last two terms in \eqref{DEM2} -- see  App.~\ref{app:M2} for more details. In \eqref{M2BPSenergy}  $K^{\underline{0}}=K^\s{M} e^{\underline{0}}_\s{M}$ is the zeroth component of $K$  in the local frame   \eqref{Made}, that is, $K^{\underline{0}}$ is the   component  of $K$ normal to $S$.

 In App.~\ref{app:M2} we prove, following \cite{Martucci:2011dn}, that the integrated forms in \eqref{M2BPSenergy} satisfy the algebraic  bound
\beq\label{M2bound}
K^{\underline{0}}\,{\rm vol}_{\cC\cap S}\geq \Omega^{\text{\tiny(M2)}}|_{\cC\cap S}\,,
\eeq
which can be regarded as a local BPS bound.\footnote{The bound \eqref{M2bound} involves top-forms on $\calc\cap S$ and should more precisely be read as a bound for the corresponding coefficients with respect to a given reference top-form, fixing the overall sign ambiguity by requiring  that $ \text{vol}_{\calc\cap S}$ is  positive.} Hence we can conclude that 
\beq\label{M2bound2}
I_{\text{\tiny(M2)}}(\epsilon)\geq 0
\eeq
and then the inclusion of fully backreacting M2-branes preserves the positivity of  BPS energy:
\beq\label{totIM2}
I(\epsilon)\geq 0\,.
\eeq
Furthermore, remembering the conclusion of  subsection \ref{sub:Mpos}, $I(\epsilon)=0$ if and only if $\epsilon$ satisfies the supersymmetry equations and  the bound \eqref{M2bound} is saturated:
\beq\label{M2susy}
K^{\underline{0}}\,{\rm vol}_{\cC\cap S}= \Omega^{\text{\tiny(M2)}}|_{\cC\cap S}\,.
\eeq
As explained in App.~\ref{app:M2}, this condition is equivalent to requiring that the M2-brane does not break the supersymmetry generated by $\epsilon$.

As in \cite{Martucci:2011dn}, the local BPS bound \eqref{M2bound} provides a  generalization to non-static settings of the ordinary calibration bounds \cite{harvey-lawson}, which instead apply only to static backgrounds. These local bounds are usually exploited to study the energetics of {\em probe} branes on fixed supersymmetric vacua. Our result \eqref{IsplitM2}  makes it clear how, even after  the brane backreaction is taken into account and  more general on-shell bulk deformations are allowed, the combination entering the local bound \eqref{M2bound} still determines the  M2 contribution $I_{\text{\tiny(M2)}}(\epsilon)$ to the positive  BPS energy $I(\epsilon)$.
Furthermore, we explicitly see how the inclusion of M2-branes cannot be represented just by their energy-momentum tensor \eqref{M2EM} (which satisfies the dominant energy condition), since this would not take into account the second (possibly negative) contribution in \eqref{M2BPSenergy}. We will explicitly show how these observations also hold for other branes in string/M-theory, and we indeed expect them to be  universal properties of supergravity models including  charged branes.

%%%%%%%%%%%%%%%%%%%%%%%%%%%%%%%%%%%%%%%%%%%%%%%%%%%%%%%%%

\subsection{Inclusion of M5-branes}
\label{sec:M5}

Conceptually, the inclusion of M5-branes should proceed as for the M2 case. In particular, we expect an M5 to contribute to the BPS energy with a non-negative contribution $I_{\text{\tiny(M5)}}(\epsilon)$ analogous to \eqref{M2BPSenergy}. This  contribution may be computed along the lines of what we did for M2-branes, starting from the M5 action of \cite{pasti-sorokin-tonin}. One technical difficulty comes from the presence of the self-dual three-form on the M5-brane, which induces an M2-charge on the M5. 

In order to alleviate the presentation, we start by simply ignoring the self-dual three-form contribution.  Then the bosonic  action associated with an M5-brane wrapping a six-dimensional submanifold $\cC$ reduces to
\beq\label{M5action}
S_{\text{\tiny(M5)}}=-2\pi\int_\cC\d^6\sigma\sqrt{-h}+2\pi\int_\cC \tilde C+\ldots \eeq
where $\tilde C$ is the dual six-form potential, locally defined by $\d\tilde C=-(*G+\frac12 C\wedge G)$, and the ellipsis refers to the omitted terms depending on  the world-volume self-dual  3-form. 
The WZ term contributes to the Bianchi identity for $G$, which becomes:
\beq
\d G=\delta^{(5)}(\cC)\,.
\eeq
Also the Einstein equations get modified, by a contribution similar to the M2-brane one appearing on the right-hand side of \eqref{M2Einst}. 
By repeating the same steps followed for the M2 case, we arrive at the following M5 contribution to BPS energy: 
\beq\label{M5I}
I_{\text{\tiny(M5)}}(\epsilon)=\frac14\int_{\cC\cap\Sigma}\left(K^{\underline{0}}\,{\rm vol}_{\cC\cap\Sigma}-\Omega^{\text{\tiny(M5)}}\right)\,,
\eeq
where 
\beq\label{OmegaM5}
\Omega^{\text{\tiny(M5)}}\equiv\bar\epsilon\,\Gamma_{(5)}\epsilon\,.
\eeq
We have again  a   calibration bound 
\beq\label{M5bound}
K^{\underline{0}}\,{\rm vol}_{\cC\cap\Sigma}\geq \Omega^{\text{\tiny(M5)}}|_{\cC\cap\Sigma}\,,
\eeq
which is saturated by the supersymmetric configuration. This implies the positivity bound
\beq\label{M5bound2}
I_{\text{\tiny(M5)}}(\epsilon)\geq 0\,,
\eeq
for the M5 contribution to the BPS energy.

Coming back to the self-dual 3-form,  it induces an M2-charge on the M5-brane. Hence,  it affects the $C$ equations of motion and contributes to the Einstein equations by further terms localized on the M5-brane. These can be straightforwardly computed starting from the action of \cite{pasti-sorokin-tonin}, but we will not do this exercise here, since in section \ref{sub:ii-branes} we will  do it for D-branes supporting  general world-volume fluxes. Since M5-branes and  D-branes are related by dualities, the results of section \ref{sub:ii-branes} are sufficient to conclude  that \eqref{M5bound2}  indeed still holds once the M5 self-dual three-form flux is taken into account, and is saturated only by supersymmetric configurations. 

In summary, the total BPS energy \eqref{genEnergy} can be written as the sum of three non-negative terms 
\beq
I(\epsilon)=I_0(\epsilon)+I_{\text{\tiny(M2)}}(\epsilon)+I_{\text{\tiny(M5)}}(\epsilon)\,, 
\eeq
where  $I_0(\epsilon)$ becomes manifestly non-negative if we impose the Witten condition \eqref{eq:w}. Furthermore $I(\epsilon)=0$ if and only if the three terms separately vanish. This can happen only if the complete configuration is supersymmetric, that is, if  $\epsilon$ is a bulk supercharge and the branes preserve it.

\subsection{Positivity and stability of \texorpdfstring{$G_2$}{G2}-compactifications}
\label{sub:G2}

We can illustrate the above general results by discussing energy positivity and  stability in M-theory compactifications on special holonomy  spaces. As we have mentioned, this stability is widely expected, but checking it is a good exercise towards attacking vacua without supersymmetry.
The energy positivity associated with these kinds of configurations has been already discussed in \cite{Dai:2003jr,Dai:2004yd}, which however focused  on the gravitational sector (plus a possible energy-momentum tensor satisfying the dominant energy condition). Our results allow us to include the effects of the $G$ flux and of the M2/M5 branes as well.

For concreteness we focus on vacua of the form $M_0=\mathbb{M}_4\times Y$, where $\mathbb{M}_4$ is four-dimensional Minkowski space and $Y$ admits  a given $G_2$-holonomy metric. (Compactifications to $\mathbb{M}_{d\neq 4}$, or on $Y$ with special holonomy $\neq G_2$, are analogous.)  
In these purely geometrical  vacua, the eleven-dimensional metric is a direct sum $\d s^2_0=\d s^2_{\mathbb{M}_4}+\d s^2_Y$,
and preserves a supercharge 
\beq\label{G2killing}
\epsilon_0=\varepsilon_0\otimes\eta_0
\eeq
where $\varepsilon_0$ is a constant four-dimensional Majorana spinor, while $\eta_0$ is the covariantly constant Majorana spinor along the $G_2$-manifold $Y$, which we choose to have unit norm $\eta_0^\dagger\eta_0=1$.  
We are interested in general deformations which asymptotically reduce to the $M_0=\mathbb{M}_4\times Y$ vacuum. On $M_0$  we can introduce some adapted coordinates $(x^\mu,y^m)$, where $y^m$ are coordinates along $Y$ and $x^\mu=(t,r,\theta,\phi)$ are standard spherical coordinates over $\mathbb{M}_4$. Asymptotically, we can use these coordinates also on the deformed backgrounds and choose the nine-dimensional space-like surface $S$ to take the form $\Sigma\times Y$, where $\Sigma$ is an asymptotically flat  three-dimensional surface of constant time $t$. Correspondingly, the nine-dimensional boundary $\del S$ appearing in the BPS energy \eqref{genEnergy} takes the form  $\del S=\del\Sigma\times Y$, where $\del\Sigma$ is the two-sphere $S^2_r$ of constant radius  $r$,   in the limit $r\rightarrow \infty$. 

We impose boundary conditions where the eleven-dimensional metric and spinor of the deformed background are asymptotically $g=g_0+\Delta g$ and $\epsilon=\epsilon_0+\Delta\epsilon$, with $\Delta g$ and  $\Delta \epsilon$ of order $\calo(r^{-1})$ and such that $\cald^{0}_\s{M}\epsilon\equiv \nabla_\s{M}\Delta\epsilon\sim \calo(r^{-2})$ (in the natural asymptotically flat frame). In order to guarantee that \eqref{genEnergy} is finite, we also assume that the variation of the spin-connection $\Delta \omega^{\s{AB}}\equiv \Delta\omega_\s{M}{}^{\s{AB}}\d x^\s{M}$ and of the field-strength $\Delta G\equiv G$ are at least of order $\calo(r^{-2})$. From the identity 
\beq\label{*Eidentity} 
*E_2=-\bar\epsilon\Gamma_{(8)}\wedge\cD\epsilon\,,
\eeq 
with $\cD\epsilon\equiv\cald_\s{M}\epsilon\,\d x^\s{M}$, we see that the integrand appearing in \eqref{genEnergy}  reduces asymptotically to 
\beq\label{asympt*E}
-\bar\epsilon_0\Gamma_{0(8)}\wedge(\cD-\cD_0)\epsilon_0+\d(\ldots)+\calo(r^{-3})\,.
\eeq 
It follows that  $I(\epsilon)$ can be written in the following form --- see also \cite[Eq.~(5.18)]{legramandi-martucci-t}:
\beq\label{MIasympt}
\begin{aligned}
I(\epsilon)=&\, \frac14\int_{\del\Sigma\times Y}*(e^\s{A}\wedge e^\s{B}\wedge K^\flat)\wedge \Delta\omega_{\s{AB}}\\
&+\frac14\int_{\del\Sigma\times Y}\left(\Omega^{\text{\tiny(M2)}}\wedge *\Delta G + \Omega^{\text{\tiny(M5)}}\wedge \Delta G  \, -2 *\Omega^{\text{\tiny(M5)}}\wedge \Delta\omega\right)\,,
\end{aligned}
\eeq
where  we have introduced the one-form $K^\flat \equiv K_\s{M}\d x^\s{M}$, obtained by lowering the index of \eqref{MKdef}, and the three-form $\Delta\omega\equiv \frac12\Delta\omega_{\s{AB}}\wedge e^\s{A}\wedge e^\s{B}$, while $\Omega^{\text{\tiny(M2)}}$ and $\Omega^{\text{\tiny(M5)}}$ are defined in \eqref{OmegaM2} and \eqref{OmegaM5} respectively.
These can be expressed in terms of the associative form $ϕ$ and the co-associative form $*_7ϕ$ on the $G_2$ manifold as:
\begin{equation}
    K^\flat = v,\, \qquad Ω^\textrm{\tiny{(M2)}} = w,\, \qquad Ω^\textrm{\tiny{(M5)}} = v∧*_7ϕ +*_4w∧ϕ,\, \qquad
    *Ω^\textrm{\tiny{(M5)}} = w∧*_7ϕ + *_4v ∧ ϕ,
\end{equation}
where $v_μ \equiv \bar{ε}γ_μ ε$, $w_{μν}\equiv \bar{ε}γ_{μν} ε$, and $ε$ is a four dimensional Majorana spinor.

Note that in \eqref{MIasympt}, the forms $K^\flat$, $\Omega^{\text{\tiny(M2)}}$ and $\Omega^{\text{\tiny(M5)}}$, as well as the Hodge-star operator,  can in fact be computed by using the  supersymmetric vacuum metric $g_0$ and supercharge $\epsilon_0$. We can then identify the first two terms appearing in the second line of \eqref{MIasympt} as measuring the asymptotic central charges associated to M2-branes stretching along $\mathbb{M}_4$  and   M5-branes wrapping internal three- and four-cycles.  Since we want to consider fluctuations around the vacuum configurations, we can further restrict the boundary conditions in such a way that these terms vanish. 

Similarly,  the three-form $\Delta\omega$ may be interpreted as measuring the flux of a KK6-brane  \cite{hull-charges}, and  $*\Omega^{\text{\tiny(M5)}}$ can be considered as a corresponding calibration --- see  section 5 of \cite{legramandi-martucci-t}.  Then the last term of \eqref{MIasympt} may be interpreted as measuring the central charge of
a KK6-branes wrapping internal four-cycles and does not contribute if we assume vacuum boundary conditions. 

On the other hand, the first term on the right-hand side of \eqref{MIasympt} can be interpreted as defining the conserved component $-K^\s{M}P_\s{M}$ of the total momentum of the system \cite{nester,hull-charges}. By using the fact that $K^\s{M}=(k^\mu_0,0)$, we see that \beq
I(\epsilon)=I(\varepsilon_0)=- k^\mu_0 P_\mu
\eeq
as in \eqref{IPrel}. Our previous discussion on the positivity of $I(\epsilon)$ then translates into a fully eleven-dimensional proof of the positivity of the energy of this class of compactifications. 

Moreover, if $I(\epsilon)=0$ for any $ε_0$, spacetime coincides with the vacuum. To see this, first observe that, just as in the pure four-dimensional case, the combination of \eqref{eq:E-square}, \eqref{eq:w} and $I(\epsilon)=0$ imply (when applied to all possible $\Sigma$) that there are four independent spinors $\epsilon$ such that ${\cal D}_\s{M} \epsilon=0$. This in turn implies that there are four independent Killing vectors $\bar\epsilon \Gamma^\s{M} \epsilon$. These coincide at infinity with the four translations of Minkowski space; so their orbits  need to be four-dimensional. Since the finite transformation generated by a Killing vector is an isometry, the metric is invariant along these four-dimensional orbits; so on the union of these orbits the metric is in fact equal to its value at infinity $\mathbb{M}_4\times Y$. 

We now use the usual logic: since $I(\epsilon)$ does not depend on $\Sigma$, it is conserved; but since it vanishes for any $\varepsilon_0$ only on the  vacuum,  the  latter is fully stable.
As usual, this conclusion assumes the existence of a solution of the Witten condition \eqref{eq:w}. In the absence of fluxes and branes, this has been rigorously proven in \cite{Dai:2003jr,Dai:2004yd}. As at the end of subsections \ref{sub:mink4} and \ref{sub:Mpos}, one could argue that $\Gamma^a\cald_a$ does not have  any normalizable  zero-mode $\epsilon_{\rm zm}$.   

Finally, note that we may in fact relax the above restrictions on the asymptotic behaviour of $G$ and $*G$ and allow for possible non-trivial asymptotic fluxes thereof. In this case the positivity of \eqref{MIasympt} would correspond to an extremality bound as in \cite{gibbons-hull,gibbons-hawking-horowitz-perry}, which would include the contribution of  four-dimensional string and membrane central charges.

%%%%%%%%%%%%%%%%%%%%%%%%%%%%%%%%%%%%%%%%%%%%%%%%%%%%%%%%%%

\subsection{Stability of \texorpdfstring{$\mathrm{AdS}_4$}{AdS4} vacua}
\label{sub:AdS4-stab}

Again in the spirit of gaining experience that might be later useful for supersymmetry-breaking vacua,
we can adapt the discussion of the previous subsection to M-theory compactifications down to AdS$_d$. 
For $d=4$, the vast majority of such supersymmetric vacua are of the famous Freund--Rubin (FR) type: 
\begin{equation}
    \d s^2_{0} = \dd s^2_{\mathrm{AdS}_4} + 4L^2\dd s^2_{Y} \, ,\qquad G_4 = \frac3L \mathrm{vol}_{\mathrm{AdS}_4}\,.
\end{equation}
The AdS radius is $L^6= \frac{N }{3\cdot 2^7\mathrm{Vol}(Y)}$, where $N$ is a positive integer; the internal space $Y$ admits an internal Killing spinor $\eta$, satisfying $(D_m -\frac \ii 2 m\gamma_m) \eta=0$; these include weak $G_2$ ($\mathcal{N}=1$), Sasaki--Einstein manifolds ($\mathcal{N}=2$), 3-Sasaki ($\mathcal{N}=3$). The total supercharge is then  $\epsilon_0=\hat\varepsilon\otimes\eta$, where $\hat\varepsilon(x)=S(x)\varepsilon_0$ is one of AdS Killing spinors mentioned after \eqref{eq:DW-4d}. We will modify these solutions in section \ref{sub:whiff} to break supersymmetry.

In order to address the stability issue, one again needs to specify appropriate  boundary conditions. The discussion in sections \ref{sub:AdS4} and \ref{sub:susy-br-4} suggests that we should require the metric at infinity to factorize as $\d s^2_4 + \d s^2_7$, with $\rho^{-2}\d s^2_4$ inducing on the boundary a metric conformal to $\mathbb{R}\times S^2$, and $\d s^2_7 = 4L^2\d s^2_{Y} + {\cal O}(\rho^{-2})$. Moreover, the spinors $\epsilon$ are such that $K^\s{M}=\bar \epsilon \Gamma^\s{M} \epsilon$ is asymptotically $(k_0^\mu=\bar\varepsilon \gamma^\mu \varepsilon,0)$.
These boundary conditions should be confirmed by uplifting the analysis in \cite{Hollands:2005wt,amsel-hertog-hollands-marolf,Hollands:2006zu} to the case with extra dimensions. In particular one should check that the BPS energy is still conserved with this choice, and that the Witten condition can be imposed; we hope to return to this in the future.  

The logic then proceeds along lines that are by now familiar. We choose a space-like surface $S$ that asymptotically takes the form $\Sigma\times Y$, with $\Sigma$ being a space-like three-dimensional slice with boundary $\del\Sigma\simeq  S^2$. The BPS energy \eqref{genEnergy} is defined by an integral over  $\del S=\del\Sigma\times Y\simeq S^2\times Y$. We may also apply \eqref{MIasympt} and provide a more explicit geometrical formula for $I(\epsilon)$, but the supersymmetry argument which led from \eqref{Qdef}  to \eqref{genEnergy} anyway guarantees that $I(\epsilon)=I(\varepsilon_0)$ must take the form \eqref{AdS4I} in terms of the ${\rm SO}(2,3)$ charges $J_{\s{AB}}$. As in subsection \ref{sub:AdS4}, this implies the positivity of $E\equiv J_{04}\geq 0$. A version of the argument in the previous subsection shows that the only zero-energy configuration is the vacuum AdS$_4\times Y$. As usual, if the energy is conserved it now follows that such a vacuum is stable.\footnote{It was found in \cite{bizon-rostworowski} that small perturbations in AdS can refocus until they collapse into a black hole; this is a different notion of stability from the one of the vacuum we consider in this paper.}

\section{Positivity and stability in type II} % (fold)
\label{sec:ii}

In this section we will show that the main points of the above M-theory discussion  hold also for ten-dimensional type II string theories, up to some technical but interesting details.

The type II fields are: the string frame metric $g_{\s{MN}}$; a two-form potential $B$ with closed three-form field-strength $H=\dd B$; RR potentials $C_k$ with field-strengths $F_k = \dd C_k - H \wedge C_{k-2}$; two Majorana--Weyl (MW) gravitinos $\psi_{\s{M}}^a$; two MW dilatinos $\lambda^a$. It will be convenient to use the \textit{democratic} formalism,  where all RR field-strengths $F_k$ are included for all $0\le k\le 10$, even in IIA and odd in IIB, and the formal total sum $F\equiv \sum_k F_k$  satisfies
\begin{equation}\label{eq:sF}
	F= * \lambda F\,.
\end{equation} 
Here $\lambda$ acts on a form by reversing the order its indices, \ie $\lambda F_k=(-)^{\frac{k(k-1)}2}F_k$, and  the Hodge-star is computed in the string frame, which is more natural for the democratic formulation. The theory is defined by a pseudo-action \cite{democratic}, whose equations of motion must be supplemented by the self-duality conditions \eqref{eq:sF}. 

The supersymmetry parameters are two MW spinors $\epsilon_a$, $a=1,2$. It is convenient to collect all the spinors into doublets: $\psi_\s{M}\equiv \binom{\psi_{1\s{M}}}{\psi_{2\s{M}}}$, and so on. The supersymmetry transformations of the fermions, with fermions set to zero, now read 
\beq\label{IIFsusy}
\delta_\epsilon\psi_\s{M}={\mathcal D}_\s{M}\epsilon\,, \qquad \delta_\epsilon\lambda={\mathcal O}\epsilon\,,
\eeq 
where\footnote{Our conventions are as in \cite{legramandi-martucci-t}. They differ from those in  \cite{lust-marchesano-martucci-tsimpis} by a sign change $H\rightarrow -H$.}
\begin{equation}\label{eq:susy-op-ii}
\begin{split}
	{\mathcal D}_\s{M}&\equiv D_\s{M} \otimes \mathbf{1}_2 -\frac14 H_\s{M} \otimes \sigma_3 + {\mathcal F}\, \Gamma_\s{M}\, ,\\
	{\mathcal O}&\equiv \dd \phi \otimes \mathbf{1}_2- \frac12 H \otimes \sigma_3 + \Gamma^\s{M} {\mathcal F} \,\Gamma_\s{M}\,,\\ &\quad\text{with}\quad 
	{\mathcal F} \equiv  \frac{\ee^\phi}{16} \left(\begin{smallmatrix}
		0 & F  \\ \pm \lambda (F) & 0 
	\end{smallmatrix}\right)\ \text{in} \, \quad \overset{\text{\tiny{IIA}}}{\text{\tiny{IIB}}}\,.
\end{split}
\end{equation}
 Here and in what follows, the upper/lower sign will refer to IIA/IIB. We work in string units $2\pi\sqrt{\alpha'}=1$.

\subsection{BPS energy} % (fold)
\label{sub:supercharge-ii}

 In order to identify the BPS energy, we will start from the supercharge generator, following the same strategy used in M-theory. The explicit form of the supercharge can be obtained by various means. For instance, it can be identified in IIA by dimensional reduction of the M-theory supercharge \eqref{Qdef} -- see App.~\ref{sec:m-ii} -- and  then  extrapolated to IIB. 
The result is
\begin{equation}\label{typeIIQ}
\begin{split}
Q(\epsilon)= \int_{\del S} \ee^{-2\phi}\,\left(\dd x^\s{M}\wedge\overline{\epsilon}\, \Gamma \Gamma_{(7)}\psi_\s{M}+\overline{\epsilon}\, \Gamma \Gamma_{(8)}\lambda\right)
	\,,
\end{split}
\end{equation}
again in the notation of \eqref{eq:G(p)}. In \eqref{typeIIQ}, $S$ is a nine-dimensional spacelike surface and we have introduced the chiral operator $\Gamma\equiv \Gamma^{\underline{01\ldots 9}}$. \footnote{\label{foot:Eframe} The supercharge \eqref{typeIIQ} admits a simpler form in terms of the Einstein frame gravitino. Denoting the Einstein frame quantities with a hat symbol $\hat{\ }$, they are related to those in the string frame by $g_{\s{MN}}=\ee^{\phi/2} \hat g_{\s{MN}}$, $\epsilon= \ee^{\phi/8}\hat\epsilon$,  $\lambda=\ee^{-\phi/8}\hat\lambda$, $\psi_\s{M}=\ee^{\phi/8}\left(\hat\psi_\s{M}+\frac18\hat\Gamma_\s{M}\hat\lambda\right)$. Then \eqref{typeIIQ} takes the form $Q(\hat\epsilon)=\int_{\del S}\dd x^\s{M}\wedge \overline{\hat\epsilon}\, \Gamma \hat\Gamma_{(7)}\hat\psi_\s{M}$. }

We can then compute  $\delta_\epsilon Q(\epsilon)=\{Q(\epsilon),Q(\epsilon)\}$ by using \eqref{IIFsusy}, and obtain in this way  the  BPS energy: 
\begin{equation}\label{eq:E2-ii}
\begin{aligned}
I(\epsilon)&= \int_{\del S} *E_2
\end{aligned}
\end{equation}
where we have introduced the Nester-like two form
\begin{equation}
E_2\equiv -\frac12\ee^{-2 \phi} \bar \epsilon (\Gamma_{\s{MN}}{}^\s{P} {\mathcal D}_\s{P}- \Gamma_{\s{MN}}{\mathcal O}) \epsilon\,\d x^{\s{M}}\wedge \d x^{\s{N}}
\,.
\end{equation}
More explicitly, we can write the  components of $E_2=\frac12 E_{\s{MN}}\dd x^\s{M} \wedge \dd x^\s{N}$ as follows:
\begin{equation}
\label{eq:E2-ii-exp}
	E_{\s{MN}}= -\ee^{-2 \phi} \bar \epsilon\, \Gamma_{\s{MN}}{}^\s{P} (D_\s{P}+{\mathcal A}_\s{P}) \epsilon\,,
	\end{equation}
	where $D_\s{M}$ is the ordinary spinor covariant derivative and  ${\mathcal A}_\s{M}$ is defined by the relation 
\beq\label{defA}	
\cald_\s{M}-\frac18\Gamma_\s{M}\calo=D_\s{M}+\cala_\s{M}\,.
\eeq	
In the following we will also need the identity
\begin{equation}\label{Adef}
	\Gamma^\s{P}{}_{\s{MN}} {\mathcal A}_\s{P} = \frac14 \Gamma_{[\s{M}} H \Gamma_{\s{N}]} \otimes \sigma_3 - \Gamma_{\s{MN}} \dd \phi -2 \Gamma_{[\s{M}} {\mathcal F}\, \Gamma_{\s{N}]}\,.
\end{equation}
This identification  can be obtained from \eqref{eq:E2-ii-exp} by using (\ref{eq:lrG0}) with $k=2$ and $H_\s{M}=\frac12 \{\Gamma_\s{M},H\}$.

% subsection supercharge-ii (end)

\subsection{Main identity} % (fold)
\label{sub:identity-ii}

Once again we need to compute the divergence of $E_2$. We begin with 
\begin{align}\label{eq:nablaE-ii-0}
	\ee^{2 \phi} &\nabla_\s{M} E^{\s{MN}}= -2 \partial_\s{M} \phi E^{\s{MN}} - \overline{D_\s{M} \epsilon}\, \Gamma^{\s{MNP}} {\mathcal A}_\s{P} \epsilon - \bar \epsilon\, D_\s{M}(\Gamma^{\s{MNP}} {\mathcal A}_\s{P} \epsilon)\\
	\nonumber
	&=-\overline{D_\s{M} \epsilon}\Gamma^{\s{MNP}} (D_\s{P} \epsilon + 2 {\mathcal A_\s{P}}\epsilon) - \bar \epsilon\, \Gamma^{\s{MNP}}(D_\s{M} D_\s{P} -2 \partial_\s{M} \phi{\mathcal A}_\s{P} + [D_\s{M},\Gamma^{\s{MNP}} {\mathcal A}_\s{P}]) \epsilon\,.
\end{align}
We have skipped a few steps similar to (\ref{eq:nablaE-0}), and we have used (\ref{eq:e'e}) for $k=2$ to derive\linebreak $\bar \epsilon\,\Gamma^{\s{MNP}} {\mathcal A}_\s{P} D_\s{M} \epsilon = \overline{D_\s{M} \epsilon}\Gamma^{\s{MNP}} ({\mathcal A}_\s{P}+ 2 \partial_\s{P} \phi)\epsilon$.

The term $\bar \epsilon\,(\cdots) D_\s{M}\epsilon$ now does not vanish automatically (unlike $A^{\s{NP}}$ in (\ref{eq:AB-m})). But by (\ref{defA}) we expect the appearance of the square of this operator. The remainder looks at first quite complicated, but a little experimentation shows that it simplifies if we subtract a further $\overline{{\mathcal O} \epsilon} \Gamma^\s{N} {\mathcal O } \epsilon$. In order to write the final result in the most transparent way, we first use the spinor $\epsilon$ to construct the future-pointing time-like or null vector
\beq
K\equiv\frac12\,\bar\epsilon\,\Gamma^\s{M}\epsilon\,\del_\s{M}
\eeq
and the (poly)forms:
\begin{subequations}
\begin{align}
&\Omega^{\text{\tiny (F1)}}\equiv \frac12\,\bar\epsilon\,\Gamma_{(1)}\otimes\sigma_3\epsilon\quad,\quad 
\Omega^{\text{\tiny (NS5)}}\equiv  \frac12\ee^{-2 \phi}\,\bar\epsilon\,\Gamma_{(5)}\otimes\sigma_3\epsilon\,,\\
&\Omega^{\text{\tiny (D)}}=\sum_{k\ \text{even/odd}}\Omega^{\text{\tiny (D)}}_{k}\equiv \sum_{k} \ee^{- \phi}\,\bar\epsilon_1\Gamma_{(k)}\epsilon_2\,.\label{Dcal}
\end{align}
\end{subequations}
Then, with some further computations presented in App.~\ref{app:ii}, we are able to show that
\begin{equation}\label{eq:nablaE-ii}
\framebox{$\begin{aligned}
	\nabla_\s{M} E^{\s{MN}} =& \ee^{-2 \phi} \overline{\left({\mathcal D}_\s{M} - \frac18 \Gamma_\s{M} {\mathcal O}\right)\epsilon}\,\Gamma^{\s{MPN}} \left({\mathcal D}_\s{P} - \frac18 \Gamma_\s{P} {\mathcal O}\right)\epsilon - \frac18 \ee^{-2 \phi} \overline{{\mathcal O} \epsilon} \,\Gamma^\s{N} {\mathcal O } \epsilon\\
	+&\cale^{\s{NP}}K_\s{P}+\frac12\calh^{\s{NP}}\Omega^{\text{\tiny (F1)}}_\s{P}-\frac12( \dd H\wedge \d x^\s{N})\cdot\Omega^{\text{\tiny (NS5)}}  +\frac12 ( \dd_H F\wedge \d x^\s{N})\cdot\Omega^{\text{\tiny (D)}}\,.
\end{aligned}$}	
\end{equation}
Here $\d_H\equiv \d -H\wedge$ and the tensors $\cale_{\s{MN}}$ and $\calh_{\s{MN}}$ are defined by
\begin{subequations}\label{eq:eom-ii}
\begin{align}
	\ee^{2 \phi}\cale_{\s{MN}}&\equiv R_{\s{MN}}-\frac12 g_{\s{MN}} R +2[(\nabla_\s{M} \nabla_\s{N} - g_{\s{MN}}\nabla^2) \phi+ g_{\s{MN}}|\dd \phi|^2]-\frac12 T^{\s{(HF)}}_{\s{MN}}\,,\label{eq:eom-iia}\\
	\calh_2& =\frac12\calh_{\s{MN}}\d x^\s{M}\wedge \d x^\s{N}\equiv *\left[\dd(\ee^{-2\phi}* H)-\frac12 (F \wedge \lambda F)_8\right]\,,\label{eq:eom-iib}
\end{align}
\end{subequations}
where $T^{\s{(HF)}}_{\s{MN}}\equiv H_\s{M} \cdot H_\s{N} -\frac12 g_{\s{MN}}|H|^2 +\frac12\ee^{2\phi} F_\s{M} \cdot F_\s{N}$.
Here we are using a notation similar to that in (\ref{eq:geom}), e.g.\  $H_\s{M}\equiv\iota_\s{M} H$ and $F_\s{M}\equiv \iota_\s{M} F$. Furthermore $(\cdots)_8$ denotes taking the eight-form part of the polyform.

We emphasize that \eqref{eq:nablaE-ii} is valid for both type II theories and holds identically for any (possibly off-shell) configuration.

\subsection{Positivity in the absence of branes} % (fold)
\label{sub:pos-ii}

As in section \ref{sub:Mpos}, we first consider the positivity of the BPS energy \eqref{eq:E2-ii} in the absence of branes (and orientifolds). In this case the Einstein and $B$-field equations of motion require that $\cale_{\s{MN}}=0$ and $\calh_{\s{MN}}=0$, respectively. Furthermore the $B$-field Bianchi identity is $\d H=0$ and the democratic RR equations of motion/Bianchi identities read $\d_H F=0$.\footnote{In order to get an on-shell configuration, these equations  must be supplemented by  the 
dilaton equation of motion $R-\frac12 |H|^2 -4 \ee^\phi \nabla^2 \ee^\phi=0$.}
Then only the first line in (\ref{eq:nablaE-ii}) survives; 
the second vanishes by the theory’s equations of motion and Bianchi identities.

If we again introduce an adapted  vielbein $e^\s{A}=(e^{\underline 0},e^a)$ such that  $e^{\underline 0}|_S=0$ and use Stokes' theorem, we can write \eqref{eq:E2-ii}  as $I_0(\epsilon)=\int_S\d*E_2=\int_S \text{vol}_S \nabla_\s{M} E^{\underline 0 M}$. By  using (\ref{eq:nablaE-ii})  we then get the following  on-shell BPS energy:
\begin{align}\label{eq:E-square-ii}
	I_0(\epsilon)&= \int_S \text{vol}_S\, \left[\left({\mathcal D}^a \epsilon-\frac18 \Gamma^a{\mathcal O}\epsilon\right)^\dagger\left({\mathcal D}_a \epsilon-\frac18 \Gamma_a {\mathcal O} \epsilon\right) + \frac18( {\mathcal O}\epsilon)^\dagger {\mathcal O}\epsilon-\left|\left(\Gamma^a {\mathcal D}_a-\frac98 {\mathcal O}\right)\epsilon\right|^2\right]\,.
\end{align}
So, the BPS energy $I_0(\epsilon)$ can be made manifestly non-negative if we can choose an $\epsilon$ satisfying the modified Witten identity:
\begin{equation}\label{eq:w-ii}
	\left(\Gamma^{a}{\mathcal D}_a-\frac98 {\mathcal O}\right)\epsilon=0\,.
\end{equation}
Hence, under this assumption we get  $I_0(\epsilon)\geq 0$. Furthermore this bound is saturated if and only if the background is supersymmetric.

Note that, as in the previous cases, the condition \eqref{eq:w-ii} can be related to a gauge-fixing condition for the gravitino \cite{hull-positivity}, which now reads $\Gamma^a\psi_a=\frac98\lambda$. This does not look like a standard transverse gauge, but only because we are working in the string frame, in which the  gravitino and dilatino kinetic terms are not diagonal.  Indeed the condition $\Gamma^a\psi_a=\frac98\lambda$ corresponds to the transverse gauge $\Gamma^a\hat\psi_a=0$ in the Einstein frame -- see footnote \ref{foot:Eframe}.  Furthermore, one can  argue for the absence of normalizable zero modes of the operator appearing in \eqref{eq:w-ii}, and then for the existence of a solution of \eqref{eq:w-ii}, by adapting the arguments proposed at the end of sections \ref{sub:mink4} and  \ref{sub:Mpos}.

% subsection pos-ii (end)

\subsection{Inclusion of localized sources} % (fold)
\label{sub:ii-branes}

As in the M-theory case, one can include branes, which can be part of  the background configuration or can contribute to the fluctuations around it. As we will see, the presence of O-planes can be similarly taken into account.   

The simplest example is provided by F1-strings, which works very similarly to the M2-brane case of section \ref{sec:M2}. In particular, the presence of an F1-string along a two-dimensional world-sheet $\calc$ modifies the $B$-field equations of motion to
\beq
\calh_2=*\delta^{(8)}(\calc)\,, 
\eeq
as well as the Einstein equations. The bottom line is that the first two terms in the second line of \eqref{eq:nablaE-ii} do not vanish anymore, but provide the following localized contribution to $I(\epsilon)=\int_S\d*E_2=\int_S \text{vol}_S \nabla_\s{M} E^{\s{\underline 0 M}}$:
\beq\label{IF1}
I_{\text{\tiny(F1)}}(\epsilon)=\frac12 \int_{\calc\cap S}\left(K^{\underline{0}}\,\text{vol}_{\calc\cap\Sigma}- \Omega^{\text{\tiny(F1)}}\right)\,.
\eeq
One can adapt the calculation of App.~\ref{app:C2} for M2-branes  to rewrite the local bound found in \cite[Eq.~(3.25)]{Martucci:2011dn} in the form
\beq\label{F1localbound}
K^{\underline{0}}\text{vol}_{\calc\cap S}\geq  \Omega^{\text{\tiny(F1)}}|_{\calc\cap S}\,.
\eeq
 This shows that \eqref{IF1} is non-negative. Furthermore, \eqref{F1localbound} is saturated if and only if the F1-string does not break the supersymmetry generated by $\epsilon$ \cite{Martucci:2011dn}. Hence, we get the bound 
 $I(\epsilon)=I_0(\epsilon)+I_{\text{\tiny(F1)}}(\epsilon)\geq 0$
for the total BPS energy, which is saturated if and only if the complete configuration is supersymmetric. For comparison with the D-brane discussion below, we observe that we can  rewrite \eqref{IF1} in the form
\beq\label{IF12}
I_{\text{\tiny(F1)}}(\epsilon)=\frac12 \int_{\calc\cap S}\Big[{ V}(K) \text{vol}_{\calc\cap S}- \Omega^{\text{\tiny(F1)}}\Big]\,,
\eeq
where we have introduced the following one-form defined along $\calc$:
\beq\label{F1form}
{V}\equiv h^{\alpha\beta}g_{\s{MN}}e_\alpha^{\underline{0}}\del_\beta X^\s{M} \d x^\s{N}\,, \text{ section of } T^*M|_\calc\,,
\eeq 
with $M$ denoting the ten-dimensional spacetime. (Recall that, according to our general notation,   $h_{\alpha\beta}$ denotes the  induced metric on the brane.) This equivalence can be understood by picking a $\calc$ adapted vielbein $e^\s{A}=(e^{\underline{\alpha}},e^{\hat a})$, with $e^{\hat a}|_\calc=0$ (and still imposing $e^{\underline{0}}|_S=0$). 

As with the discussion for M5-branes in section \ref{sec:M5},  the incorporation of D-branes and NS5-branes is complicated by the presence of additional world-volume fluxes, but these branes are still expected to provide non-negative localized contributions $I_{\text{\tiny(D)}}\geq 0$ and $I_{\text{\tiny(NS5)}}\geq 0$ respectively. As a preliminary easier check, let us consider an NS5-brane along a submanifold $\calc$ and neglect again its world-volume flux, as we did for M5-branes in section \ref{sec:M5}.  The Einstein equations  and the $B$-field Bianchi identity get modified -- for instance, the latter becomes $\d H=\delta^{(4)}(\calc)$ -- and the first and third term of the second line of  \eqref{eq:nablaE-ii} produce a non-negative contribution to the BPS energy of the form
\beq\label{INS5}
I_{\text{\tiny(NS5)}}(\epsilon)=\frac12 \int_{\calc\cap S}\left(e^{-2\phi}K^{\s{\underline{0}}}\,\text{vol}_{\calc\cap S}- \Omega^{\text{\tiny(NS5)}}\right)\geq 0\,,
\eeq
which vanishes if and only if the (fluxless) NS5-branes preserve the bulk supersymmetry generated by $\epsilon$. 

In order to better understand the possible role of world-volume fluxes, we now discuss in more detail the presence of D-branes.  Since they are related  to NS5-branes and M5-branes by dualities, our conclusions will provide a  convincing support to some of our previous claims.

Consider the inclusion  of a D$p$-brane. In the democratic formulation, it  modifies the RR Bianchi identities/equations of motion as follows: 
\beq\label{RRD}
\d_H F=-\lambda\left[\delta^{(9-p)}(\calc)\right]\wedge\ee^{-\calf}
\eeq
where $\calc$ now denotes the D$p$-brane world-volume  and $\calf$ is the gauge invariant world-volume field strength, such that $\d\calf=-H|_\calc$. Furthermore, by using the D-brane effective action
\beq
S_{\rm \s{D}}=-2\pi\int_\calc\ee^{-\phi}\sqrt{-\det\calm}+2\pi\int_\calc C\wedge\ee^{\calf}
\eeq
where  $\calm_{\alpha\beta}\equiv h_{\alpha\beta}+\calf_{\alpha\beta}$ and $C\equiv\sum_k C_k$, one can check that the Einstein equations and the $B$-field equations of motion get modified into 
\begin{subequations}
\label{GHD}
\begin{align}
\cale^{\s{MN}}&= \frac14e^{-\phi}\sqrt{-\det\calm_{\underline{\alpha\beta}}}\,\calm^{(\underline{\alpha\beta})}e^{\s{M}}_{\underline{\alpha}}e^{\s{N}}_{\underline{\beta}}*[e^{\underline{0\ldots p}}\wedge\delta^{(9-p)}(\calc)]\,,\\
\calh^{\s{MN}}&=-\frac12e^{-\phi}\sqrt{-\det\calm_{\underline{\alpha\beta}}}\,\calm^{[\underline{\alpha\beta}]}e^{\s{M}}_{\underline{\alpha}}e^{\s{N}}_{\underline{\beta}}*[e^{\underline{0\ldots p}}\wedge\delta^{(9-p)}(\calc)]\,,
\end{align}
\end{subequations}
where we are using the adapted vielbein $e^\s{A}=(e^{\underline\alpha},e^{\hat a})$ such that $e^{\hat a}|_{\calc}=0$. 

As in \cite{Martucci:2011dn}, it is  useful to combine $K$ and $\Omega^{\text{\tiny (F1)}}$ into the `generalized vector' 
\beq
\calk\equiv K+ \Omega^{\text{\tiny (F1)}}\,, \text{ section of } TM\oplus T^*M\,.
\eeq
Along  $\calc$, we can also introduce another generalized vector field:
\beq\label{Dgenvec}
\calv\equiv\ee^{-\phi}\frac{\sqrt{-\det\calm}}{\sqrt{-\det h}}\left(\calm^{(\alpha\beta)}\del_\alpha X^\s{M}\, g_{\s{MN}}\d x^\s{N}+\calm^{[\alpha\beta]}\del_\alpha X^\s{M}\,\del_\s{M}\right)e^{\underline{0}}_\beta\,, 
\eeq
which is a section of $(TM\oplus T^*M)|_\calc$ and can be regarded as the D-brane counterpart of \eqref{F1form}.
By using \eqref{RRD} and \eqref{GHD}, we see that the first, second and fourth terms in the second line of \eqref{eq:nablaE-ii} are non-vanishing in presence of  the  D$p$-brane, which then  contributes to $I(\epsilon)=\int_S \text{vol}_S \nabla_\s{M} E^{\s{\underline 0 M}}$  by
\beq\label{Dpcont} 
\begin{aligned}
 I_{\text{\tiny(D)}}(\epsilon)=\frac12 \int_{\calc\cap S}\Big[\calv(\calk)\,\d\text{vol}_{\calc\cap S}- \Omega^{\text{\tiny(D)}}\wedge\ee^{\calf}\Big]\,,
\end{aligned}
\eeq
where $\calv(\calk)$ denotes the natural pairing between the generalized vectors $\calv$ and $\calk$:
\beq
\calv(\calk)=\ee^{-\phi}\frac{\sqrt{-\det\calm}}{\sqrt{-\det h}}\left(\calm^{(\alpha\beta)} K_\s{M} \del_\alpha X^\s{M}+\calm^{[\alpha\beta]}\Omega^{\text{\tiny (F1)}}_\s{M}\del_\alpha X^\s{M}\right)e^{\s{\underline{0}}}_\beta\,.
\eeq
Note the formal analogy between \eqref{Dpcont} and \eqref{IF12}.
Now, adapting again the steps of App.~\ref{app:C2}, one can rewrite the local bound (4.33) of \cite{Martucci:2011dn} in the form
\beq\label{Dlocbound}
\calv(\calk)\,\d\text{vol}_{\calc\cap S}\geq \left(\Omega^{\text{\tiny(D)}}\wedge\ee^{\calf}\big|_{\calc\cap S}\right)_{\rm top}\,,
\eeq
which is saturated precisely if the D-brane preserves $\epsilon$. 
Hence, as expected, the complete  D-brane contribution \eqref{Dpcont} is always non-negative and vanishes only if the D-brane preserves $\epsilon$. By duality, this strongly supports our claim that these properties hold also for the complete $I_{\text{\tiny(NS5)}}$ (and $I_{\text{\tiny(M5)}}$). 

We then conclude that, if we choose an $\epsilon$ satisfying \eqref{eq:w-ii}, the total BPS energy 
\beq\label{Itotii}
I(\epsilon)=I_0(\epsilon)+ I_{\text{\tiny(F1)}}(\epsilon)+ I_{\text{\tiny(D)}}(\epsilon)+ I_{\text{\tiny(NS5)}}(\epsilon)
\eeq 
satisfies the bound
\beq
I(\epsilon)\geq 0\,,
\eeq
which is saturated if and only if   $\epsilon$ satisfies the bulk supersymmetry equations and all the branes preserve $\epsilon$,  that is, if and only if  the complete configuration is supersymmetric.

We would now like to consider the possible presence of  orientifolds. Suppose that   $\iota:M\rightarrow M$ denotes the orientifold involution associated with an O$p$-plane, located at the corresponding fixed locus $\calc$. The O$p$-plane contributes to the effective action as a D$p$-brane with a tension/charge rescaled by a factor $-2^{p-5}$ and a vanishing $\calf$, which is consistent with the fact that the $B$-field is odd under the orientifold involution and then has vanishing pull-back on the O$p$-plane. So, the O$p$-plane contribution $I_{\text{\tiny (O)}}$ to the BPS energy takes the form \eqref{Dpcont} with $\calf=0$, up to an overall $-2^{p-5}$ factor. Furthermore, the orientifold projection on the supersymmetry generator $\epsilon\equiv \binom{\epsilon_1}{\epsilon_{2}}$ is
\beq\label{spinproj}
\iota^*\epsilon_2=\epsilon_1\,,\qquad \iota^*\epsilon_1=(-)^{\lfloor\frac{p-1}{2}\rfloor}\epsilon_2\,,
\eeq
see for instance  \cite{koerber-tsimpis}. Hence, from \eqref{Dcal} we see that $\Omega^{\text{\tiny(D)}}$ is invariant under the orientifold involution:
\beq
\iota^*\Omega^{\text{\tiny(D)}}=\Omega^{\text{\tiny(D)}}\,.
\eeq
Furthermore, since the O$p$-plane world-volume $\calc$ is fixed under the orientifold involution, compatibility with  \eqref{spinproj} requires  that along the O$p$-plane one must impose  the boundary condition
\beq
(\epsilon_1-\Gamma_{\underline{0\ldots p}}\epsilon_2)|_\calc=0\,,
\eeq
where we are again using an  adapted frame $e^\s{A}=(e^{\underline{\alpha}},e^{\hat a})$ with $e^{\hat a}|_\calc=0$.  This condition is precisely equivalent to requiring that the local bound \eqref{Dlocbound} (with $\calf=0$) is saturated along the O$p$-plane. Hence $I_{\text{\tiny (O)}}(\epsilon)\equiv 0$ and the BPS energy $I(\epsilon)$ is still given by the non-negative combination \eqref{Itotii}.

We close this section by emphasising that so far we have  neglected higher order curvature corrections to the brane effective actions, which would modify the form of the corresponding localized BPS energy.\footnote{World-volume higher curvature corrections play for instance and key role in the D-brane mediated instabilities recently discussed in \cite{Marchesano:2021ycx}.} For instance, the curvature corrections of \cite{green-harvey-moore,cheung-yin,minasian-moore} would modify the second term appearing in the r.h.s.\ of \eqref{Dpcont} into something of the form $-\frac12\int_{\calc\cap S}\Omega^{\text{\tiny(D)}}\wedge\ee^{\calf}\wedge (\text{curvature}) $.  The expected local brane supersymmetry suggests that $\calv$ is also modified, so that $I_{\text{\tiny (D)}}(\epsilon)\geq 0$ still holds and is saturated precisely if the D-brane preserves $\epsilon$. It would be interesting to better investigate this point, also in connection with  bulk higher-derivative corrections.

% subsection ii-branes (end)
% section ii (end)

% subsection m-branes (end)

% section m (end)

\section{Supersymmetry breaking} % (fold)
\label{sec:susy-br}

In the previous sections, we clarified the mechanism that makes supersymmetric vacua stable. We will now try to use this knowledge to provide stability arguments for vacua that break supersymmetry. Our attempts will fail to exhibit stable non-supersymmetric vacua, but fail in interesting ways that might point the way to better future strategies.

\subsection{An operator in M-theory} % (fold)
\label{sub:sbr-Mop}

The idea is a bit similar to that in section \ref{sub:susy-br-4}. A supersymmetric vacuum is defined by having a solution $\epsilon$ to ${\mathcal D}_\s{M} \epsilon =0$ in M-theory and ${\mathcal D}_\s{M} \epsilon = {\mathcal O} \epsilon=0$ in type II (recalling (\ref{McalDdef}), (\ref{eq:susy-op-ii})). We would like to find a modification ${\mathcal D}'_\s{M}$, ${\mathcal O}'$ of these supersymmetry operators such that 
\begin{itemize}
	\item [i)] an $\epsilon$ annihilated by them exists  on some non-supersymmetric vacuum, and
	\item [ii)] the BPS energy $I'(\epsilon)$ defined via the analogues of (\ref{eq:ME2}), (\ref{eq:E2-ii}) is still positive and related to the physical energy. 
\end{itemize}

This is a hard set of requirements to satisfy, but let us try anyway. There are infinitely many operators ${\mathcal D}'_\s{M}$ that one might consider, but the problem becomes more tractable if we demand only objects with a single derivative to appear. As reviewed in App.~\ref{app:M-susy-eom}, the consistency conditions $[{\mathcal D}_\s{M}, {\mathcal D}_\s{N}]\epsilon=0$ for supersymmetry imply most of the equations of motion. Having more than one derivative in the modified ${\mathcal D}'_M$ would imply more than two derivatives in the consistency conditions $[{\mathcal D}_\s{M}', {\mathcal D}_\s{N}']\epsilon=0$, and it might be hard to make these in turn compatible with the equations of motion. 
Another restriction that seems reasonable is that, in order to avoid gauge dependence, only the field-strengths appear and not the potentials.

In M-theory, the most general allowed operator obtained by deforming $D_\s{M}$ would now be\footnote{One additional term we could add here is $a_4 Γ_\s{M} D$, with $D= Γ^\s{N} D_\s{N}$ the Dirac operator. Consider $ Γ^\s{M} {\mathcal D}'_\s{M} \epsilon = 0$ with this modification; for $a_4≠-1/11$, this equation determines $D \epsilon$, which can be fed back into ${\cal D}'_\s{M} \epsilon = 0$ to obtain a new equation without the new term. If $a_4 = -1/11$, ${\cal D}'_\s{M}$ contains the \emph{conformal Killing operator} $D_\s{M} - \frac1d Γ_\s{M} D$. We will ignore this particular case in what follows.}
\begin{equation}\label{eq:D'}
	{\mathcal D}'_\s{M} = D_\s{M}+\frac1{24}(a_1 \Gamma_\s{M} G + a_2 G \Gamma_\s{M}) + a_3 \Gamma_\s{M}\,,
\end{equation} 
where the $a_i$ are constants. Comparing with (the second expression in) (\ref{McalDdef}), we see that the supersymmetric case is recovered for $a_1=-1$, $a_2=3$, $a_3=0$.

Notice that in particular we are not considering a modification of the type 
\begin{equation}\label{eq:DV}
	{\mathcal D}'_\s{M} = {\mathcal D}_\s{M} + V_\s{M}\,.
\end{equation} 
This would have the interesting feature that the two-form $(E'_2)_\s{MN}= - \bar \epsilon \Gamma_{\s{MN}}{}^\s{P} {\mathcal D}'_\s{P} \epsilon$ defined with this operator is in fact equal to the $E_2$ in (\ref{eq:ME2}); so the positivity argument would still apply.  Moreover solving ${\mathcal D}'_\s{M} \epsilon =0$ would be superficially reminiscent of $(D_m + \ii A_m) \eta_+=0$, which does have a solution on a K\"ahler manifold (although with a crucial difference of an $\ii$). However, such a vector $V_\s{M}$ cannot be an external piece of data: for the positivity argument to apply, it has to be defined for all possible configurations with a certain boundary condition. So it has to be somehow defined in terms of the fields. Given the requirements we listed at the beginning of this section, there does not seem to be any such vector field, and hence we will ignore the possibility (\ref{eq:DV}). 

% subsection sbr-Mop (end)

\subsection{Skew-whiffed and Englert vacua} % (fold)
\label{sub:whiff}

We will first look at requirement i) above: namely, whether we can solve ${\mathcal D}'_\s{M} \epsilon=0$ on any non-supersymmetric vacua.

\textit{Skew-whiffed} vacua are obtained from a supersymmetric FR vacuum (section \ref{sub:AdS4-stab}) by reversing the orientation of the internal space, or in other words by mapping $G\to -G$ \cite{duff-nilsson-pope-skew,duff-nilsson-pope}.  As reviewed in section \ref{sub:AdS4-stab},  in a FR solution AdS$_4\times Y$ supersymmetry requires solving the internal Killing spinor equation $(D_m -\frac \ii 2 m\gamma_m) \eta=0$ on $Y$. Reversing the sign of $G$ flips the sign of $m$, and the new spinor equation now has no solution (except when $Y=S^7$); so supersymmetry is broken. But the equations of motion are still satisfied because they are quadratic in $G$. As noted already in \cite{duff-nilsson-pope-skew,duff-nilsson-pope}, these solutions are automatically stable under small perturbations.\footnote{Further analysis of stability of these solutions under squashing was carried out in \cite{ekhammar-nilsson,nilsson-padellaro-pope}.} 
On these non-supersymmetric solutions, we can clearly solve ${\mathcal D}_\s{M}' \epsilon=0$ with
\begin{equation}\label{eq:skew}
	a_1=1 \, ,\qquad a_2=-3 \, ,\qquad a_3=0\,;
\end{equation}
in other words, with the operator obtained from the supersymmetric ${\mathcal D}_\s{M}$ by reversing the sign of $G$. 

We next consider the \textit{Englert} vacua \cite{englert-susybr}. Here, $Y$ is a \textit{weak $G_2$} manifold, namely one with a $G_2$-structure $\phi$ such that\footnote{Another class of solutions uses instead the ${\rm SU}(3)$-structure on a Sasaki--Einstein manifold \cite{pope-warner-susybr,bobev-halmagyi-pilch-warner,pilch-yoo}. It would be interesting to apply our methods to this case as well; we thank N.~Bobev for the suggestion.
} 
\begin{equation}
	\d \phi = - 4 *_7 \phi\,.
\end{equation}
This implies that the cone over $Y$ has $\mathrm{Spin}(7)$ holonomy, so $Y$ admits a Killing spinor $\eta_0$ with $m=1$. As a consequence, $Y$ is also Einstein: more specifically $R_{mn}= 6 g_{mn}$. Taking the fields as 
\begin{equation}\label{eq:englert}
	\d s^2_{11} = L^2 \d s^2_{\mathrm{AdS}_4} + r_0^2 \d s^2_{Y}\, ,\qquad
	 G= g_0 \mathrm{vol}_{\mathrm{AdS}_4} + g_1 *_7 \phi \,,
\end{equation}
the equations of motion have two branches of solutions: one with $g_1=0$, leading back to FR, and one with 
\begin{equation}\label{eq:englert2}
	g_0 = \frac9{25}r_0^3 \, ,\qquad g_1^2= 4 r_0^6 \, ,\qquad L= \sqrt{\frac3 {10}} r_0\,.
\end{equation}
A perturbative instability was found for these vacua in \cite{page-pope-stab-englert}, but only when $Y$ has more than one Killing spinor. One can check from \eqref{eq:englert}
that the brane nucleation condition \eqref{eq:QT} is \textit{not} satisfied for the easiest case of M2-branes (as noted in \cite{suh-ads4} for some particular cases).

A solution of ${\mathcal D}_\s{M}' \epsilon=0$ is found as $\epsilon= \zeta_+ \otimes  \eta + \zeta_- \otimes  \eta^\mathrm{c}$. $\zeta_-=(\zeta_+)^\mathrm{c}$, with ${}^\mathrm{c}$ denoting Majorana conjugation; as usual we assume the Killing spinor equation $D_\mu \zeta_\pm= \frac12 \gamma_\mu \zeta_\mp$ also along AdS$_4$; and $\eta = \ee^{\ii \alpha} \eta_0$, with $\alpha$ a constant and $\eta_0$ Killing. With these assumptions, both the internal and external components of our equation become algebraic, and can be solved for
\begin{equation}\label{eq:englert3}
	a_2= \frac{3\pm\sqrt{114}}{10} \, ,\qquad a_1 = 3 -a_2 \, ,\qquad a_3 = \pm\frac{21-2a_2}3 \, ,\qquad \tan2 \alpha =\frac{3-2a_2}{4a_2}	\,;
\end{equation}
the two signs are independent. 

% subsection whiff (end)

\subsection{Lack of positivity} % (fold)
\label{sub:susybr-pos}

Having found some examples of solutions to ${\mathcal D}_\s{M}'  \epsilon= 0$, we now look at requirement ii) above, namely that one can still prove positivity and stability with the modified operator ${\mathcal D}_M'$. 

In principle there are various terms that can ruin positivity. In (\ref{eq:nablaE-0}), all the terms beyond the first should now be re-examined. Already the second term  $\bar \epsilon\, A^{\s{NP}} D_\s{P} \epsilon$ is worrisome: it is unlikely to have a fixed sign because of the derivative of $\epsilon$, so it had better vanish.\footnote{One might want to take care of this term by ``completing the square'' to reabsorb it in the term $\overline{{\cal D}'_M \epsilon} \Gamma^{MPN} {\cal D}_P \epsilon$; but this requires looking for a $X_M$ such that $X_M Γ^{MNP} = [G, Γ^{NP}]$; there appears to be no natural choice of such an object.} Repeating the steps in (\ref{eq:ANP=0}) we see that in our present more general setting 
$2A^{\s{NP}}=-3(3a_1+a_2)[\Gamma^{\s{NP}},G]$.
In particular this imposes
\begin{equation}\label{eq:a2=3a1}
    a_2 = - 3a_1\,.
\end{equation}
We see from (\ref{eq:skew}), (\ref{eq:englert3}) that the  skew-whiffed vacua do satisfy this, while the Englert vacua do not. While it is conceivable that a positivity property might be proven by more ingenious methods without assuming (\ref{eq:a2=3a1}), in the rest of this subsection we will assume that it must hold. 

With this assumption, defining $E'_{\s{MN}}= - \bar \epsilon \Gamma_{\s{MN}}{}^\s{P} {\mathcal D}'_\s{P} \epsilon$ similar to (\ref{eq:ME2}), (\ref{eq:nablaE}) is modified to
\begin{equation}\label{eq:nablaE-susybr}
\begin{split}
	\nabla_\s{M} E'^{\s{MN}} &= \overline{{\mathcal D}'_\s{M} \epsilon} \Gamma^{\s{MPN}} {\mathcal D}'_\s{P} \epsilon +\frac14(2\calg^{\s{NP}}- a_1^2 T^{\s{NP}}_{\s{(G)}})K_\s{P}\\
	&+\frac14 \bar \epsilon\left[\dd x^\s{N} \wedge \left(a_1\d G -a_1\d* G+\frac12 a_1^2G\wedge G - 12 a_1 a_3 G -360 a_3^2\right) \right]\epsilon\,,
\end{split}
\end{equation}
where $\calg_{\s{MN}}\equiv R_{\s{MN}}-\frac12 g_{\s{MN}}R$ is the usual Einstein tensor; see App.~\ref{app:M-susybr} for the details of this computation.
In the absence of branes, the terms multiplying $K_\s{P}$ can be reassembled using the equations of motion as $(1-a_1^2)T^{\s{NP}}_{\s{(G)}}$.  By using \eqref{genEnergy2} we see that, if we assume $|a_1|\leq 1$, the contribution of this term  to $I'(\epsilon)$ is positive because the flux stress-energy tensor obeys the dominant energy condition.

Of more concern are the terms on the second line of (\ref{eq:nablaE-susybr}). We can use the equations of motion (\ref{eq:Geom}) to get rid of $\d G$ and $\d *G$, turning the quadratic flux term into $\frac12 a_1(a_1+1) G \wedge G $. However, there is no reason to believe this term to have a definite sign for all solutions.\footnote{Consider for example a solution where $K$ is timelike; here $\epsilon$ will define an $\mathrm{SU}(5)$-structure $(J_2, \Omega_5)$ in the remaining directions (studied in \cite{gauntlett-pakis}), in terms of which the term of interest is proportional to $ J_2 \wedge G \wedge G$, a top-form in $d=10$. Choosing a holomorphic vielbein $h^a$, in terms of which $J_2=\frac\ii2\sum_a h^a \wedge \overline{h^a}$, it is easy to see that the forms $G= 2 {\rm Re}(h^1 \wedge h^2 \wedge h^3 \wedge \overline{h^4})$ and $G=J_2\wedge J_2$ result in opposite signs for $ J_2 \wedge G \wedge G$.} Furthermore, if $a_3\neq 0$ we also have a linear $G$ term, which is even more clearly of indefinite sign.

It is also interesting to consider the effect of localized sources.  The same steps leading to \eqref{M2BPSenergy} now give a localized contribution to the BPS-energy
\beq\label{M2BPSenergy2}
I'_{\text{\tiny(M2)}}(\epsilon)= \frac14\int_{\cC\cap S}\left(K^{\underline{0}}\,{\rm vol}_{\cC\cap\Sigma}-|a_1|\Omega^{\text{\tiny(M2)}}\right)
\eeq
where the overall sign of the term including $\Omega^{\text{\tiny(M2)}}$ has been fixed without loss of generality, since it can be changed   by inverting the M2 orientation, that is, by swapping the M2 for an anti-M2 brane, or viceversa. 

The sign of $\int_{\cC\cap S}\Omega^{\text{\tiny(M2)}}$ depends on the  M2 orientation.
 Clearly $I'_{\text{\tiny(M2)}}(\epsilon)$ is manifestly positive if $\int_{\cC\cap S}\Omega^{\text{\tiny(M2)}}\leq 0$. If instead $\int_{\cC\cap S}\Omega^{\text{\tiny(M2)}}> 0$,  the second term on the r.h.s.\ of \eqref{M2BPSenergy2} is negative. However, we can still use the algebraic local bound \eqref{M2bound}, which implies that
\beq
I'_{\text{\tiny(M2)}}(\epsilon)\geq \frac14(1-|a_1|)\int_{\cC\cap S}\Omega^{\text{\tiny(M2)}}\,.
\eeq
 We then see that  $I'_{\text{\tiny(M2)}}(\epsilon)$ has a definite sign if $|a_1|\leq 1$, which is the same condition that we found  above. This suggests that  vacua can be unstable under M2 nucleation precisely if $|a_1|> 1$. Note that  this conclusion does not apply to the skew-whiffed case (in which $a_1=1$).    
A similar argument could be repeated for the non-supersymmetric generalization of the M5 contribution \eqref{M5I}. However, in this case the neglected world-volume fluxes could affect the final conclusion.  

We saw earlier that the Englert vacua don't satisfy (\ref{eq:a2=3a1}). Even if we ignore this issue and repeat (\ref{eq:nablaE-susybr}) with $a_2\neq -3 a_1$, it turns out that the sign of the $T^{\s{NP}}_{\s{(G)}}$ is negative; so this class of solutions is unlikely to enjoy a positivity theorem by the spinorial strategy we are considering.

The skew-whiffed vacua seem more promising in this respect, but they still have a non-vanishing $ G \wedge G $ term, which as we saw earlier does not have a definite sign. So we again conclude that no positivity theorem exists with the present spinorial method. This does not prove that these vacua are unstable, but certainly seems to give evidence in that direction.

% subsection susybr-pos (end)

\subsection{Type II} % (fold)
\label{sub:susybr-ii}

The supersymmetry transformations in M-theory already contain all the possible terms that one can write with at most one derivative, except for one term; this led us to (\ref{eq:D'}). 

In contrast, in the type II operators (\ref{eq:susy-op-ii}) it would be possible to add several new terms. For example one could change the $2\times 2$ matrices acting on the index $a$ of the spinor doublet $\epsilon_a$. Or one could add entirely new terms, such as $H \Gamma_\s{M}$, $\Gamma_\s{M} {\mathcal F}$, $\partial_\s{M} \phi$ to ${\mathcal D}_\s{M}$ or ${\mathcal F}$ to ${\mathcal O}$, tensored by any $2\times 2$ matrix compatible with the chirality of the spinors. Even more drastically one could violate the democratic structure of the operator and add a term only involving a single RR-form degree. This gives rise to a bewildering array of possibilities, that we will not explore here. For example it would be interesting to try to reproduce for example the GKP supersymmetry breaking \cite{giddings-kachru-polchinski}, or the supersymmetry-breaking mechanisms discussed in \cite{lust-marchesano-martucci-tsimpis,legramandi-t}, from this point of view.

Instead of attempting a general analysis, it would be reasonable to add one such term at a time. The simplest possibilities would be
\begin{equation}\label{eq:D'-II}
    {\mathcal D}'_\s{M} = {\mathcal D}_\s{M} + \Gamma_\s{M} \otimes a\,,\qquad {\mathcal O}' = {\mathcal O}+ 1\otimes b\,,
\end{equation}
where $a$, $b$ are some $2\times 2$ matrices. Notice however that this only makes sense in IIA provided $a$ and $b$ are off-diagonal, since in IIB ${\mathcal D}_\s{M} \epsilon$ and $\Gamma_\s{M}\otimes a \epsilon$ have opposite chiralities, and so do ${\mathcal O} \epsilon$ and $1 \otimes  b \epsilon$. 
It would be interesting to apply to \eqref{eq:D'-II} in IIA the same program outlined earlier in this section for M-theory.  

% subsection susybr-ii (end)

% section susy-br (end)

\section*{Acknowledgements}

We would like to thank N.~Bobev, D.~Cassani, B.~De Luca, G.~Dibitetto, C.~Hull, G.~Lo Monaco and B.~Nilsson for useful discussions and correspondence. 
We are supported in part by INFN and by MIUR-PRIN contract 2017CC72MK003.

\appendix

%%%%%%%%%%%%%%%%%%%%%%%%%%%%%%%%%%%%%%%%%%%%%%%%%%%%%%%%%%%%%%%%%%%%%%%%%%%%%%%%%%%

\section{Useful spinorial identities} % (fold)
\label{app:pre}

Here we recall some standard definitions and techniques useful in any number $d$ of dimensions; we assume Lorentz signature.

The easiest gamma matrix identities are
\begin{equation}\label{eq:lrG0}
	\begin{split}
	\Gamma_\s{M} \Gamma^{\s{N_1\cdots N_k}} &= \Gamma_\s{M}{}^{\s{N_1\cdots N_k}}+ k \delta_{\s{M}}^{\s{[N_1}} \Gamma^{\s{\cdots N_k]}} \, ,\\
	\Gamma^{\s{N_1\cdots N_k}}\Gamma_\s{M} &= (-1)^k(\Gamma_\s{M}{}^{\s{N_1\cdots N_k}}- k \delta_{\s{M}}^{\s{[N_1}} \Gamma^{\s{\cdots N_k]}}) \, .
	\end{split}
\end{equation}
Viewing $\Gamma^{\s{N_1\cdots N_k}}$ as the image of the Clifford map ${}_\slash$
\begin{equation}\label{eq:clifford}
	[\dd x^{\s{M_1}} \wedge \cdots \wedge \dd x^{\s{M_k}}]_{\slash} = \Gamma^{\s{M_1\cdots M_k}}\,,
\end{equation} 
equations \eqref{eq:lrG0} can be thought of as operator identities:
\begin{equation}\label{eq:lrG}
	\stackrel\to {\Gamma^\s{M}} =  g^{\s{MN}}\iota_\s{N}+\dd x^\s{M} \wedge 
	\, ,\qquad
	\stackrel\leftarrow {\Gamma^\s{M}} = (-g^{\s{MN}}\iota_\s{N}+\dd x^\s{M}\wedge)(-1)^\mathrm{deg} \,.
\end{equation}
The arrows denote action from the left and right: $ {\Gamma^\s{M}} \alpha_{\slash} = (\stackrel\to {\Gamma^\s{M}} \!\!\alpha)_{\slash}$, 
$ \alpha_{\slash}\Gamma^\s{M} = (\stackrel\leftarrow {\Gamma^\s{M}}\!\! \alpha)_{\slash}$. Moreover $\deg \alpha_k \equiv k \alpha_k$, with the subscript denoting form degree; $\iota_\s{M} (\dd x^{\s{N_1}}\wedge\cdots \wedge \dd x^{\s{N_k}})\equiv k \delta_{\s{M}}^{\s{[N_1}} \dd x^{\s{N_2}} \wedge \cdots \wedge \dd x^{\s{N_k]}}$ is the contraction operator.  

A useful identity following from (\ref{eq:lrG})  is
\begin{equation}\label{eq:gag}
	\Gamma_\s{M} \alpha_k \Gamma^\s{M} = (-1)^k (d-2k)\alpha_k\,.
\end{equation}
One particular consequence of this and (\ref{eq:lrG0}) for $k=2$ is
\begin{equation}\label{eq:g1g3}
	\Gamma_\s{M} \Gamma^{\s{MNP}} = \Gamma^{\s{MNP}} \Gamma_\s{M} = (d-2) \Gamma^{\s{NP}}\,.
\end{equation}
Sometimes, in order to avoid ambiguities, we underline flat indices. For instance, the chiral matrix is $\Gamma= c \Gamma^{\underline{01\ldots d-1}}$. In general $c$ is chosen such that $\Gamma^2=1$, and in both $d=10$ and 11 we can just take $c=1$. Under the Clifford map (\ref{eq:clifford}), left multiplication by $\Gamma$ is related to the Hodge star: 
\begin{equation}\label{eq:G*}
	\stackrel\to{\Gamma} = c * \lambda\,,
\end{equation}
where $\lambda \alpha_k \equiv (-1)^{\lfloor k/2 \rfloor} \alpha_k$. Our Hodge-star operator $*$ is defined by
\beq\label{def*}
*(e^{\s{A_1}}\wedge \ldots \wedge e^{\s{A_k}})=\frac{1}{(d-k)!}\epsilon_{\s{B_1\cdots B_{d-k}}}{}^{\s{A_1\cdots A_k}}e^{\s{B_1}}\wedge \cdots \wedge e^{\s{B_{d-k}}}
\eeq
where $e^{\s{A}}=e^{\s{A}}_{\s{M}}\d x^\s{M}$ is a vielbein and  the totally antisymmetric  $\epsilon_{\s{A_1\cdots A_d}}$ is such that $\epsilon_{\underline{012\cdots}}= 1$. 
Notice that  $(* \lambda)^2 = -(-)^{\lfloor d/2 \rfloor}$ in Lorentzian signature. Using this and (\ref{eq:lrG}) we also obtain that the Hodge operator exchanges wedges with contractions:
\begin{equation}\label{eq:*dx}
	* \lambda\,  \dd x^\s{M} \wedge = (-1)^{d-1}\iota^\s{M} * \lambda \, ,\qquad
	* \lambda\,  \iota^{\s{M}}  = (-1)^{d-1}\dd x^\s{M} \wedge * \lambda \,.
\end{equation}
In the slightly different notation introduced in (\ref{eq:G(p)}), we can also show 
\begin{equation}\label{eq:*G(p)}
	\Gamma\Gamma_{(k)}=c (-1)^{\lfloor d/2 \rfloor} *\lambda(\Gamma_{(d-k)})\,.
\end{equation}
The natural inner product and norm-squared of forms are:
\beq
\alpha_k\cdot \beta_k\equiv \frac1{k!}\alpha_{\s{M_1\cdots M_k}}\beta^{\s{M_1\cdots M_k}}\,,\qquad |\alpha_k|^2\equiv \alpha_k\cdot \alpha_k\,.
\eeq
There is also a natural inner product among bispinors, related under (\ref{eq:clifford}) to the one among forms:
\begin{equation}
	\mathrm{Tr}((\lambda \alpha_k)_{\slash} (\beta_k)_{\slash}) = 2^{\lfloor d/2 \rfloor} \alpha_k \cdot \beta_k\, ;\qquad (\lambda \alpha_k)_{\slash}= (-1)^k \Gamma_{\underline{0}} (\alpha_k{}_{\slash})^\dagger \Gamma^{\underline{0}}\,.
\end{equation}
 Repeated application of (\ref{eq:lrG}) gives the other useful identities
\begin{subequations}
\begin{align}
	\label{eq:ii-tr}
	(\iota^\s{M} \alpha_k)\cdot (\iota^\s{N} \beta_k) -\frac12 \alpha_k \cdot \beta_k\, g^{\s{MN}} &=\frac{(-1)^{k+1}}{2 \cdot 2^{\lfloor d/2 \rfloor}}  \mathrm{Tr}\left((\lambda \alpha_k)_{\slash} \Gamma^{\s{(M}} \slashed{\beta}_k \Gamma^{\s{N)}}\right)\,,\\
	\label{eq:idx-tr}
	(\iota^\s{M} \alpha_k)\cdot (\dd x^\s{N} \wedge \alpha_k)  &=\frac{(-1)^{k}}{2 \cdot 2^{\lfloor d/2 \rfloor}}  \mathrm{Tr}\left((\lambda \alpha_k)_{\slash} \Gamma^{\s{[M}} \slashed{\alpha}_k \Gamma^{\s{N]}}\right)\,.
\end{align}	
\end{subequations}
Notice that the left-hand side of (\ref{eq:ii-tr}) has the form of a stress-energy tensor associated to a $k$-form field-strength.

We will sometimes use Fierz identities. These come about by expanding a bispinor along the $\Gamma^{\s{N_1\cdots N_k}}$. When $d=$even, the set of such objects $k= 0,\,\cdots,\,d$ is a basis; when $d=\mathrm{odd}$, there are redundancies relating $k$ to $d-k$, so we can expand along only $k=0,\,\cdots,\,\left(d-1\right)/2$. 

The coefficients of such expansions often involve inner products $\bar\epsilon \epsilon'$, where  \begin{equation}
	\bar \epsilon \equiv \epsilon^\dagger \Gamma_{\underline0}\,.
\end{equation} 
We also recall the intertwiner property $\Gamma_{\s{M}}^\dagger \Gamma^{\underline0} = - \Gamma^{\underline 0} \Gamma_\s{M}$. In both $d=10$ and $11$, one can work in a basis where all gamma matrices are real; a Majorana spinor is then such that $\epsilon^* = \epsilon$, and for two Majorana spinors one then easily sees
\begin{equation}\label{eq:e'e}
	\bar \epsilon'\, \Gamma_{\s{M_1 \cdots M_k}} \epsilon = (-1)^{\lfloor (k-1)/2 \rfloor} \bar \epsilon\, \Gamma_{\s{M_1 \cdots M_k}} \epsilon'  \,.
\end{equation}
It is also useful to note that 
\begin{equation}\label{eq:Gepsbar}
	\overline{ \Gamma_{\s{M_1 \cdots M_k}} \epsilon} = (-1)^k\bar \epsilon\,\Gamma_{\s{M_k \cdots M_1}} = (-1)^{\lfloor (k-1)/2\rfloor}\bar\epsilon\,\Gamma_{\s{M_1 \cdots M_k}}\,.
\end{equation}
The Clifford map (\ref{eq:clifford}) also works well at the differential level. One can extend the spinorial covariant derivative to bispinors $C$ as $D_\s{M} C \equiv \partial_\s{M} C + \frac14\omega_{\s{M}}^{\s{AB}}[ \Gamma_{\s{AB}},C]$; this is then related to the ordinary bosonic covariant derivative $\nabla_\s{M}$ by
\begin{equation}\label{eq:Dnabla}
	D_\s{M} (\alpha_{\slash})= (\nabla_\s{M} \alpha)_{\slash}\,.
\end{equation}

%%%%%%%%%%%%%%%%%%%%%%%%%%%%%%%%%%%%%%%%%%%%%%%%%%%%%%%%%%%%%%%%%%%%%%%%%%%%%%%%%%%

\section{Some details on M-theory} % (fold)
\label{app:details}

The two derivative action of 11 dimensional supergravity reads:
\begin{equation}\label{eq:M-action}
\begin{split}
    S = \frac{2\pi}{\lp^9}\bigintssss \Big( R *1 &- 2\ii \overline{ψ} ∧ Γ_{(8)} ∧ Dψ - \frac12 G ∧*G - \frac{1}{6}G∧G∧C \\
    & -\frac{i}{6} \overline{ψ} ∧ Γ_{(8)} ∧ \left( Γ_{(1)}\slashed{G} - 3 \left(\iota_{(1)} G\right)_\slash \right)ψ  + \mathcal{O}\left(ψ^4\right) \Big),
\end{split}
\end{equation}
where $D$ is the spin covariant derivative, and $G=\d C$, and $\slashed{G}$ and $\left(\iota_\s{N}G\right)_\slash$ are defined below \eqref{McalDdef} in agreement with the general \eqref{eq:clifford}.
The supersymmetry variation of the $3$-form gauge field and the gravitini are given by:
\begin{equation}
    δ_ϵ C_{\s{MNP}} = -3 \bar{ϵ}\, Γ_{\s{\left[MN\right.}}ψ_{\s{\left. P \right]}}, \qquad
    δ_ϵ ψ_\s{M} = \mathcal{D}_\s{M} ϵ  \,,
\end{equation}
where $\mathcal{D}_\s{M}$ is defined in \eqref{McalDdef}.

\subsection{Supercharge from the Noether theorem}
\label{sub:noether}

Let us now briefly summarize a derivation of the supercurrent following an infinitesimal variation under local supersymmetry, along the lines of \cite{hristov-toldo-vandoren}.

Given a Lagrangian $L\left(X,∂ X\right)$ depending on fields $X$ and their derivatives $∂_\s{M} X$, its variation is given by
\begin{equation}
\begin{aligned}
    δL &=\sum_X \left( \frac{δL}{δX} δX + \frac{δL}{δ∂_\s{M}X} ∂_\s{M} δX \right)\\
    &=\sum_X  \left( \frac{δL}{δX} δX - \left(∂_\s{M} \frac{δL}{δ∂_\s{M}X}\right)δX \right)
    +\sum_X ∂_\s{M} \left( \frac{δL}{δ∂_\s{M} X}δX \right) \\
    &\equiv \sum_X \left( \mathcal{E}_X\, δX +∂_\s{M} N^\s{M} \right),
\end{aligned}
\end{equation}
where $\mathcal{E}_X$ represents the equations of motion for $X$.
Now, the variation of the Lagrangian under an infinitesimal supersymmetry transformation gives $δ_ϵ L \equiv ∂_\s{M} V^\s{M}$. Identifying this with the above gives
\begin{equation}
    ∂_M \underbrace{\left( V^\s{M} - N^\s{M} \right)}_{\equiv J^\s{M}} =\sum_X \mathcal{E}_X\, δX \overset{\textrm{on-shell}}{=} 0,
\end{equation}
which gives the supercurrent $J^\s{M}= V^\s{M} - N^\s{M}$ that is conserved on shell. The corresponding supercharge is $Q = \int \d^{d-1} x J^0$, which also generates the supersymmetry transformations of the fields: $δ_ϵ X = \left\{ Q,X \right\}$.

% subsection rev (end)

% subsection pre (end)

\subsection{Noether charge in M-theory} % (fold)
\label{app:noether}

Applying to M-theory the Noether procedure outlined in \ref{sub:noether} we get,
\begin{equation}
    N^\s{M} = \frac{δL}{δ\left( ∂_\s{M} ψ_\s{N} \right)}δ_ϵψ_\s{N} + \frac{1}{3!}\frac{δL}{δ\left( ∂_\s{M} C_{\s{NPQ}} \right)}δ_ϵC_{\s{NPQ}} + \cdots,
\end{equation}
where the ellipsis represents variations with respect to the vielbien and the spin connection. These eventually cancel out and so we will not write them here.
Up to an overall rescaling by $\lp^9/(8\pi)$, this gives
\begin{equation}\label{eq:Nm-m}
\begin{aligned}
    N^\s{M} &=  -\frac{\ii e}{2}\overline{ψ}_\s{P} Γ^{\s{PMN}} \mathcal{D}_\s{N} ϵ + \frac{3e}{4\cdot 4!} G^{\s{MNPQ}}\bar{ϵ}\, Γ_{[\s{NP}}ψ_{\s{Q}]}\\
    &+ \frac{1}{4\cdot 12³}ϵ^{\s{MNPQ M_5…M_{11}}}G_{\s{M_5…M_8}}C_{\s{M_9 M_{10} M_{11}}}\bar{ϵ}\, Γ_{[\s{NP}}ψ_{\s{Q}]} + \cdots,
\end{aligned}
\end{equation}
where again the ellipsis represents, in addition to the variation with respect to $e_\s{M}^\s{A}$ and $ω$ above, quadratic terms in $ψ$.
The surface terms arising from an infinitesimal variation of the action under the local supersymmetry transformation give $δ_ϵ L = \frac{8\pi}{\lp^9} ∂_\s{M} V^\s{M}$, where
\begin{equation}
\begin{aligned}
    V^\s{M} &=  \frac{\ii  e}{2}\overline{ψ}_\s{P} Γ^{\s{PMN}} \mathcal{D}_\s{N} ϵ + \frac{3e}{4\cdot 4!} G^{\s{MNPQ}}\bar{ϵ}\, Γ_{[\s{NP}}ψ_{\s{Q}]}\\
    &+ \frac{1}{4\cdot 12³}ϵ^{\s{MNPQ M_5…M_{11}}}G_{\s{M_5…M_8}}C_{\s{M_9 M_{10} M_{11}}}\bar{ϵ}\, Γ_{[\s{NP}}ψ_{\s{Q}]} + \cdots,
\end{aligned}
\end{equation}
and the ellipsis has the same meaning as above.
Together with \eqref{eq:Nm-m}, this gives the supercurrent
\begin{equation}\label{eq:Jm-m}
    J^\s{M} = V^\s{M} - N^\s{M} = \ii e \overline{ψ}_\s{P} Γ^{\s{PMN}} \mathcal{D}_\s{N} ϵ.
\end{equation}
For our purposes, it is convenient to use the fermionic real supercurrent obtained from \eqref{eq:Jm-m} by  considering $\epsilon$ to be commuting and dropping the overall $\ii$.
The corresponding fermionic supercharge  $Q(\epsilon)$ is then given by the integral:
\begin{equation}\label{eq:Qm-m}
    Q = \int_S \overline{ψ}_\s{P} Γ^{\s{PMN}}\mathcal{D}_\s{N} ϵ * \d x_\s{M}
    = \int_{∂S} \overline{ψ}_\s{P} Γ^{\s{P}}{}_{\s{MN}} ϵ * \d x^{\s{MN}}
    = -\int_{∂S} \overline{ϵ}\, Γ_{(8)} ∧ ψ\,.
\end{equation}
In the second step, we have used the gravitino equations of motion $Γ^{\s{MNP}} \cald_\s{N} {ψ}_\s{P}=0+\calo(\psi^3)$.

% subsection noether (end)

\subsection{The divergence identity} % (fold)
\label{app:m}
We will now give some more details about the computations leading to (\ref{eq:nablaE}). For simplicity in the rest of this appendix, as in most of the paper, we mostly drop the Clifford map symbol ${}_\slash$, whose implicit presence should be clear from the context.  

As a warm-up we first note that, from (\ref{eq:gag}):
\begin{equation}\label{eq:GammaG}
	\Gamma^\s{M} G \Gamma_\s{M} = 3 G \, ,\qquad
	\Gamma^\s{M} (- \Gamma_\s{M} G + 3 G \Gamma_\s{M}) = (-11 + 3\cdot3)G = -2 G\,.
\end{equation}
We then evaluate the commutator term in (\ref{eq:nablaE-0}):
\begin{equation}\label{eq:lin}
	\begin{split}
		\Gamma^{\s{NMP}}&[D_\s{M}, -\Gamma_\s{P} G + 3 G \Gamma_\s{P}]\buildrel{(\ref{eq:Dnabla})}\over= \Gamma^{\s{NMP}}(-\Gamma_\s{P} \nabla_\s{M} G + 3 \nabla_\s{M} G \Gamma_\s{P})\\
		&\buildrel{(\ref{eq:g1g3}), (\ref{eq:lrG0}), (\ref{eq:GammaG})}\over=6\Gamma^{\s{[M}} \nabla_\s{M} G \Gamma^{\s{N]}}
		\buildrel{(\ref{eq:lrG})}\over=6[(\dd x \wedge + \iota)(\dd x \wedge - \iota)]^{\s{[MN]}} \nabla_\s{M} G\\
		&= 6(\dd x^\s{M} \wedge \dd x^\s{N} \wedge - \iota^\s{M}\iota^\s{N}) \nabla_\s{M} G
		\buildrel{(\ref{eq:*dx})}\over= 6(-\dd x^\s{N} \wedge \dd G + \iota^\s{N} * \dd * G)\,.
	\end{split}
\end{equation}
We now look at the terms in (\ref{eq:nablaE-0}) that are quadratic in $G$.
Using (\ref{eq:lrG0}) for $k=2$, it is now easy to evaluate $A^{NP}$ in (\ref{eq:AB-m}):
\begin{align}\label{eq:ANP=0}
		\nonumber
		A^{\s{NP}}=&(-G \Gamma_\s{M} + 3 \Gamma_\s{M} G) (\Gamma^\s{M}\Gamma^{\s{NP}}-2 g^{\s{M[N}}\Gamma^{\s{P]}})\\
		&- (\Gamma^{\s{NP}} \Gamma^\s{M}- 2 \Gamma^{\s{[N}} g^{\s{P]M}})(-\Gamma_\s{M} G + 3 G \Gamma_\s{M}) \\
		\nonumber=&-2 G \Gamma^{\s{NP}} -2(-G \Gamma^{\s{[N}} +3 \Gamma^{\s{[N}} G) \Gamma^{\s{P]}}
		+ 2 \Gamma^{\s{NP}} G +2 \Gamma^{\s{[N}}(- \Gamma^{\s{P]}} G + 3 G \Gamma^{\s{P]}})=0\,. 
\end{align}

Another useful consequence of (\ref{eq:lrG0}) is:
\begin{equation}
	\Gamma^{\s{MNP}}= \Gamma^\s{M} \Gamma^\s{N} \Gamma^\s{P} - g^{\s{MN}} \Gamma^\s{P} - \Gamma^\s{M} g^{\s{NP}} + \Gamma^\s{N} g^{\s{MP}}\,.
\end{equation}
Using this and (\ref{eq:GammaG}), we also get
\begin{equation}\label{eq:BN-0}
	Q^\s{N} = 9 (G \Gamma^\s{N} G + \Gamma_\s{M} G \Gamma^\s{N} G \Gamma^\s{M})\,.
\end{equation}
For our purposes it is now useful to apply the Fierz identity, \ie to expand (\ref{eq:BN-0}) in the basis $\Gamma^{\s{N_1\cdots N_k}}$, $k=0,\,\cdots,\,\left(d-1\right)/2=5$. This works out to
\begin{equation}\label{eq:BN-1}
	Q^\s{N} = \sum_{k=0}^5 \frac1{32 k!} \mathrm{Tr}(Q^\s{N} \Gamma_{\s{N_k \cdots N_1}}) \Gamma^{\s{N_1 \cdots N_k}}\,.
\end{equation}
 Actually, it is easy to see that $\Gamma_{\underline0} (Q^\s{N})^\dagger \Gamma^{\underline0}= - Q^\s{N}$, and $\Gamma_{\underline0}(\Gamma^{\s{N_1\cdots N_k}})^\dagger \Gamma^{\underline0}= (-1)^{\lfloor k+1/2 \rfloor} \Gamma^{\s{N_1\cdots N_k}}$; so in fact only $k=1,2,5$ appear in (\ref{eq:BN-1}). The traces can now be simplified using (\ref{eq:gag}) again. The $k=1$ and $k=2$ terms give the quadratic $G$ contributions to the equations of motion (\ref{eq:geom}), (\ref{eq:Geom}), while in $k=5$ the two terms cancel each other:
\begin{equation}\label{eq:Qfierz}
\begin{split}
	Q^\s{N} &= \frac1{32} \left(\mathrm{Tr}(Q^\s{N} \Gamma^\s{P}) \Gamma_\s{P} + \frac12\mathrm{Tr}(Q^\s{N} \Gamma^{\s{QP}} ) \Gamma_{\s{PQ}}\right)\\
	&=36 \left(T^{\s{NP}}_{\s{(G)}}\Gamma_\s{P} - \frac1{4(4!)^2} \epsilon^{\s{NPQ M_1 \cdots M_8}} G_{\s{M_1 \cdot M_4}} G_{\s{M_5 \cdot M_8}} \Gamma^{\s{PQ}}\right)\\
	&= 36 T^{\s{NP}}_{\s{(G)}}\Gamma_\s{P} - 18 \iota^\s{N} * (G \wedge G)\,.
\end{split}
\end{equation}
Thus as promised, this gives the quadratic terms in the equations of motion in (\ref{eq:nablaE}).

\subsection{The supersymmetry-breaking case} % (fold)
\label{app:M-susybr}

Here we will show briefly how the computation is modified for the operator (\ref{eq:D'}) with $a_2=-3a_1$, namely 
\begin{equation}
	{\mathcal D}'_\s{M} = D_\s{M}+\frac{a_1}{24}( \Gamma_\s{M} G -3 G \Gamma_\s{M}) + a_3 \Gamma_\s{M}\,.
\end{equation}
Following the same steps as in (\ref{eq:nablaE-0}), the explicit $G$ terms there are rescaled by $-a_1$; (\ref{eq:AB-m}) are replaced by
\begin{equation}
		A'{}^{\s{NP}} = -a_1 A^{\s{NP}} + 24 a_3 (\Gamma_\s{M} \Gamma^{\s{MNP}}- \Gamma^{\s{MNP}}\Gamma_\s{M}) \buildrel{(\ref{eq:ANP=0}), (\ref{eq:g1g3})}\over=0
\end{equation}
and $Q'{}^{\s{N}}$ given by
\begin{align}
    \nonumber
	\frac{Q'{}^\s{N} +a_1 Q^\s{N}}{24a_3} &=  
	-a_1\Gamma_\s{M} \Gamma^{\s{MNP}}( -\Gamma_\s{P} G + 3 G \Gamma_\s{P})- a_1 (-G \Gamma_\s{M} + 3 \Gamma_\s{M} G) \Gamma^{\s{MNP}} \Gamma_\s{P} + 24 a_3 \Gamma_\s{M} \Gamma^{\s{MNP}} \Gamma_\s{P} \\
	\nonumber
	&\buildrel{(\ref{eq:g1g3})}\over= -9 a_1 (\Gamma^\s{N} \Gamma^\s{P} - g^{\s{NP}})( -\Gamma_\s{P} G + 3 G \Gamma_\s{P}) 
	-9 a_1 (-G \Gamma_\s{M} + 3 \Gamma_\s{M} G)(\Gamma^\s{M} \Gamma^\s{N} - g^{\s{MN}}) \\ 
	&\qquad + 90 \cdot 24 a_3 \Gamma^\s{N}
	\buildrel{(\ref{eq:GammaG})}\over= 36 a_1 \{ \Gamma^\s{N}, G\} +  90 \cdot 24 a_3 \Gamma^\s{N}
	\\
	\nonumber
	&\buildrel{(\ref{eq:lrG})}\over= 
	12(3 a_1  \d x^\s{N} \wedge G +  180 a_3 \d x^N )_{\slash}\,.
\end{align}
Putting it all together, one arrives at (\ref{eq:nablaE-susybr}).

\subsection{Equations of motion from supersymmetry} % (fold)
\label{app:M-susy-eom}

Finally we show here how the Einstein equations are (almost completely) implied by supersymmetry and the $G$ equations. This was shown in \cite[App.~B]{gauntlett-pakis}, but some results we saw in this appendix allow us to derive it more quickly. This is useful for us in the context of modifying ${\mathcal D}_\s{M}$ to break supersymmetry, in section \ref{sec:susy-br}.

We consider the operator 
\begin{equation}
	\Gamma^\s{M} [{\mathcal D}_\s{M}, {\mathcal D}_\s{N}]\,.
\end{equation}
The purely gravitational term $\Gamma^\s{M} [D_\s{M},D_\s{N}]$ is well-known to be $\frac12 R_{\s{NP}}\Gamma^\s{P}$; this is how a manifold with a covariantly constant spinor in Euclidean signature is shown to be Ricci-flat. The remaining terms can again be separated into those involving $G$ linearly, and those involving it quadratically.

The linear term proceeds similar to (\ref{eq:lin}):
\begin{equation}\label{eq:lin-eom}
	\begin{split}
		&\Gamma^\s{M} [D_{\s{[M}}, -\Gamma_{\s{N]}} G + 3 G \Gamma_{\s{N]}}]
		= -\Gamma^\s{M} \Gamma_{\s{[N}} \nabla_{\s{M]}} G + 3 \Gamma^\s{M} \nabla_{\s{[M}} G \Gamma_{\s{N]}} \\
		&\buildrel{(\ref{eq:gag})}\over= \frac12 \Gamma_\s{N} \Gamma^\s{M} \nabla_\s{M} G+ \frac32 \Gamma^\s{M} \nabla_\s{M} G \Gamma_\s{N} \buildrel{(\ref{eq:lrG})}\over=
		2(-\d x_\s{N} \wedge + 2 \iota_\s{N}) (\d G + * \d\! * G)\,.
	\end{split}
\end{equation}
For the quadratic term, we first use (\ref{eq:GammaG}) repeatedly to obtain:
\begin{equation}
	\Gamma^\s{M}[ - \Gamma_\s{M} G + 3 G \Gamma_\s{M}, - \Gamma_\s{N} G + 3 G \Gamma_\s{N}]
	= -9 (G \Gamma_\s{N} G + \Gamma^\s{M} G \Gamma_\s{N} G \Gamma_\s{M}) + 9 \Gamma_\s{N} G^2 -3 \Gamma_\s{N} \Gamma^\s{M} G^2 \Gamma_\s{M} \,.
\end{equation}
The parenthesis on the right-hand side is just $Q^\s{N}$ from (\ref{eq:BN-0}), so we can evaluate it as in (\ref{eq:Qfierz}). For the other two terms, we first notice that $G^2=\frac12 \{G,G\}= (G^2)_0 + (G^2)_4 + (G^2)_8$: it only has zero-, four- and eight-form parts. From (\ref{eq:gag}) we then have
\begin{equation}
	3G^2- \Gamma^\s{M} G^2 \Gamma_\s{M} = -8 (G^2)_0 -8 (G^2)_8 =-8 (|G|^2 + G \wedge G)\,.
\end{equation}
Recalling now the first of (\ref{eq:lrG0}) and putting everything together, we arrive at
\begin{equation}\label{eq:susy->eom}
	\Gamma^\s{M} [{\mathcal D}_\s{M} , {\mathcal D}_\s{N}]= \frac12 {\mathcal E}_{\s{NP}}\Gamma^\s{P} +\frac1{12} (\d x_\s{N} \wedge - 2 \iota_\s{N}) \left(\d G + * \left(\d * G +\frac12 G \wedge G\right)\right)\,,
\end{equation}
where recall that ${\mathcal E}_{NP}=0$ is the Einstein equation of motion from (\ref{eq:geom}). 

The argument is now standard, and we only repeat it here for completeness. If the $G$ Bianchi identity and equations of motion hold (away from M2 and M5 branes), acting with (\ref{eq:susy->eom}) on a supercharge $\epsilon$ we get ${\mathcal E}_{\s{NP}} \Gamma^\s{P} \epsilon =0$. For a Majorana spinor $\epsilon$ in $d=11$, the bilinear $K_\s{M}=\bar \epsilon \Gamma_\s{M} \epsilon$ is either time-like or null. If it is timelike, $\epsilon$ has no one-form that annihilates it; so all components of ${\mathcal E}_{\s{NP}}$ are zero. If $K_\s{M}$ is null, it is the only one-form that annihilates $\epsilon$. In an adapted vielbein such that $K= e^+$, all components of ${\mathcal E}_{\s{NP}}$  except ${\mathcal E}_{\s{N-}}$ are zero; since ${\mathcal E}$ is symmetric, in fact all components except ${\mathcal E}_{--}$ are zero.\footnote{If there is at least another supercharge with a different $K$ (which is the case for vacuum compactifications with a $d\ge 3$ external spacetime) then all components ${\mathcal E}_{\s{NP}}=0$.}

% subsection M-susy-eom (end)

%%%%%%%%%%%%%%%%%%%%%%%%%%%%%%%%%%%%%%%%%%%%%%%%
%\subsection{Calibration}
%\label{app:calibration1}

%%%%%%%%%%%%%%%%%%%%%%%%%%%%%%%%%%%%%%%%%%%%%%%%%%%%%%%%%%%%%%%%%%%%%%%

\section{Inclusion of M2-branes}
\label{app:M2}

In this appendix we provide details of the derivation of \eqref{M2BPSenergy} and of the bound \eqref{M2bound}. 

\subsection{M2 BPS-energy}

We first derive the second term appearing on the r.h.s.\ of \eqref{M2BPSenergy}, which comes from the contribution of the last term of \eqref{DEM2} inside \eqref{genEnergy2}:
\beq
\begin{aligned}
& \frac{1}4 \int_S *\d x_\s{N}\, \big(*\big[\d x^\s{N}\wedge \delta^{(8)}(\cC)\big]\big)\cdot \Omega^{\text{\tiny (M2)}}=-\frac{1}4 \int_S * e^{\s{\underline 0}}\, \big(*\big[e^{\s{\underline 0}}\wedge \delta^{(8)}(\cC)\big]\big)\cdot \Omega^{\text{\tiny (M2)}}\\
&=-\frac{1}4 \int_S {\rm dvol}_S\, \big(*_{S} \delta^{(8)}(\cC)|_\Sigma\big)\cdot \Omega^{\text{\tiny (M2)}}
= -\frac{1}4 \int_S \Omega^{\text{\tiny (M2)}}\wedge \delta^{(8)}(\cC)=-\frac{1}4 \int_{\cC\cap S} \Omega^{\text{\tiny (M2)}}\,,
\end{aligned}
\eeq
where in the second line  we have used the adapted vielbein $e^\s{A}=(e^{\s{\underline 0}},e^a)$ introduced in \eqref{Made}. Notice that instead the final result does not depend on such a choice.

In order to obtain the first term on the r.h.s.\ of \eqref{M2BPSenergy}, we first derive the formula \eqref{M2EM} of $T^{\s{MN}}_{\text{\tiny (M2)}}$. This  can be extracted from the variation of the M2 action \eqref{M2action} under a metric deformation 
\beq
\begin{aligned}
&\frac12 \int \delta g_{\s{MN}}T^{\s{MN}}_{\text{\tiny (M2)}}*1\equiv \frac{1}{2\pi}\delta S_{\text{\tiny(M2)}}= -\frac{1}{2}\int_\calc\d^3\sigma\sqrt{-h}h^{\alpha\beta}\del_\alpha X^\s{M}\del_\beta X^\s{N}\delta g_{\s{MN}}\\
& =-\frac{1}{2}\int_\calc e^{\underline{012}}\,\eta^{\underline{\alpha\beta}}e_{\underline{\alpha}}(X^\s{M})e_{\underline{\beta}}(X^\s{N})\delta g_{\s{MN}}
=-\frac{1}{2}\int\delta g_{\s{MN}} e_{\underline{\alpha}}^\s{M}e_{\underline{\beta}}^\s{N}\eta^{\underline{\alpha\beta}}\,e^{\underline{012}}\wedge \delta^{(8)}(\calc)\\
&=\frac{1}{2}\int\delta g_{\s{MN}} e_{\underline{\alpha}}^\s{M}e_{\underline{\beta}}^\s{N}\eta^{\underline{\alpha\beta}}\,*[e^{\underline{012}}\wedge \delta^{(8)}(\calc)]*1\, \,,
\end{aligned}
\eeq
where we have used the adapted vielbein \eqref{M2vielbein}. 

By using \eqref{M2EM} in \eqref{DEM2}, we see that it contributes to \eqref{genEnergy2} by the term
\beq
\begin{aligned}
&\frac14 \int K_\s{M} e_{\underline{\alpha}}^\s{M}e_{\underline{\beta}}^\s{N}\eta^{\underline{\alpha\beta}}\,*[e^{\underline{012}}\wedge \delta^{(8)}(\calc)]*\d x_\s{N}=\frac14 \int K_{\underline{\alpha}}*e^{\underline{\alpha}}\,*[e^{\underline{012}}\wedge \delta^{(8)}(\calc)] \\
&=\frac14\int K^{\underline{\alpha}}\,\iota_{\underline{\alpha}}\,[e^{\underline{012}}\wedge \delta^{(8)}(\calc)]=\frac14\int K^{\underline{0}}\,e^{\underline{12}}\wedge \delta^{(8)}(\calc)=\frac14\int_{\cC\cap S} K^{\underline{0}}\,{\rm vol}_{\cC\cap S}\,,
\end{aligned}
\eeq
where we have used a further adapted vielbein combining the properties of \eqref{Made} and \eqref{M2vielbein}. 
This completes our derivation of the M2 BPS energy \eqref{M2BPSenergy}.

%%%%%%%%%%%%%%%%%%%%%%%%%%%%%%%%%%

\subsection{M2 bound}
\label{app:C2}

Consider a spacetime foliated by space-like leaves $S$ parametrized by a `time' $t$. By using adapted coordinates  $x^\s{M}=(t,x^m)$, the line element can be written as
\beq
\begin{aligned}
\d s^2&=\ee^{2D}\d t^2+\hat g_{mn}(\d x^m+\hat V^{m}\d t)(\d x^n+\hat V^n\d t)\,,
\end{aligned}
\eeq
where $\hat g_{mn}$ is the metric induced on $S$. The adapted vielbein \eqref{Made} is then
\beq
e^{\underline{0}}=\ee^{D}\d t\,,\qquad e^a=\hat e^a_m(\d x^m+\hat V^{m}\d t)\,,
\eeq
where $\hat e^a=\hat e^a_m\d x^m$ is a vielbein on $S$. Note also that the dual frame $e_\s{A}=(e_{\underline{0}},e_a)$  is such that 
\beq 
e_{\underline 0}=\ee^{-D}(\del_t-\hat V^m\del_m)\,,\qquad e_{a}=\hat e_a\,.
\eeq
Now consider an M2-brane  world-volume $\cC$. One can impose a partial static gauge and use $t$ as time coordinate along $\cC$. The world-volume coordinates $\sigma^{\alpha}=(t,\sigma^i)$ induce a decomposition of the world-volume line-element 
\beq
\begin{aligned}
\d s^2_\cC&=\ee^{2D}\d t^2+\hat h_{ij}(\d\sigma^i+\hat  v^i\d t)(\d\sigma^j+\hat v^j\d t)\,,
\end{aligned}
\eeq
where $\hat h_{ij}$ is the induced metric on $\cC\cap S$. Furthermore, we can pick an adapted bulk vielbein $e^\s{A}=(e^{\underline{0}},e^a)=(e^{\underline\alpha},e^{\tilde a})$ satisfying the properties of \eqref{Made} and \eqref{M2vielbein}. Hence, the pull-back of $ e^{\underline\alpha}$ to $
\calc$ gives a vielbein ${\mathfrak e}^{\underline\alpha}$ of  
$\d s^2_\cC$. Note also that on $\cC$ we can identify the push-forward of the dual frame ${\mathfrak e}_{\underline\alpha}$ with the bulk $e_{\underline\alpha}$.

By following \cite{Martucci:2011dn}, we can introduce the world-volume momentum density
\beq
\calp^\s{M}=-\sqrt{-h}\,h^{t\alpha}\del_\alpha X^\s{M}\,.
\eeq
Notice that $\calp^\s{M}$ and $\bar\chi\Gamma^\s{M}\chi$ are both causal and future-pointing, for any spinor $\chi$. Hence 
\beq 
-\calp^\s{M} \bar\chi\Gamma_M\chi\geq 0\,,
\eeq
where the inequality is saturated if and only if $\chi=0$. By choosing $\chi=(\mathbf{1}-\Gamma_{\text{\tiny M2}} )\epsilon$ with
\beq
\Gamma_{\text{\tiny M2}}\equiv \frac{\epsilon^{\alpha\beta\gamma}}{
3!\sqrt{-h}}\del_\alpha X^\s{M}\del_\beta X^\s{N}\del_\gamma X^P\Gamma_\s{MNP}\,,
\eeq
and recalling that $K^\s{M}=\bar\epsilon\Gamma^\s{M}\epsilon$, we get the inequality
\beq\label{ineq1}
-\calp^\s{M} K_\s{M}\geq  -\calp^\s{M} \bar\epsilon\Gamma_\s{M}\Gamma_{\text{\tiny M2}}\epsilon\,.
\eeq
This is saturated iff $\Gamma_{\text{\tiny M2}}\epsilon=\epsilon$. 

By using the adapted vielbein above, we can write 
\beq
\begin{aligned}
-\calp^\s{M} \bar\epsilon\Gamma_\s{M}\Gamma_{\text{\tiny M2}}\epsilon&=-\frac1{3!}\calp^\s{M} e_\s{M}^{\underline\alpha}\,\epsilon^{\underline{\beta\gamma\delta}}\,\bar\epsilon\Gamma_{\underline\alpha}\Gamma_{\underline{\beta\gamma\delta}}\epsilon=-\frac12\calp^\s{M} e_\s{M}^{\underline\alpha}\eta_{\underline{\alpha\beta}}\, \epsilon^{\underline{\beta\gamma\delta}}\, \bar\epsilon\Gamma_{\underline{\gamma\delta}}\epsilon\\
&=-\frac12\calp^\s{M} e_\s{M}^{\underline\alpha}\eta_{\underline{\alpha\beta}}\, \epsilon^{\underline{\beta\gamma\delta}}\, \Omega^{\text{\tiny (M2)}}_{\underline{\gamma\delta}}=-\frac1{2\sqrt{-h}} \calp^M e^{\underline\alpha}_M\eta_{\underline{\alpha\beta}}{\mathfrak e}^{\underline\beta}_\beta \epsilon^{\beta\gamma\delta}\Omega^{\text{\tiny (M2)}}_{\gamma\delta}\\
&=\frac12 h^{t\alpha}h_{\alpha\beta}\epsilon^{\beta\gamma\delta}\Omega^{\text{\tiny (M2)}}_{\gamma\delta}=\frac12\epsilon^{t\alpha\beta}\Omega^{\text{\tiny (M2)}}_{\beta\gamma}\,,
\end{aligned}
\eeq
and then
\beq
-\calp^\s{M} \bar\epsilon\Gamma_\s{M}\Gamma_{\text{\tiny M2}}\epsilon\,\d\sigma^1\wedge\d\sigma^2=\Omega^{\text{\tiny (M2)}}|_{\cC\cap S}\,.
\eeq
Hence, with respect to the oriented two-form $\d^2\sigma\equiv \d\sigma^1\wedge\d\sigma^2>0$ along $\cC\cap S$, we can write \eqref{ineq1} in the form
\beq\label{ineq2}
-\calp^\s{M} K_\s{M}\d^2\sigma\geq \Omega^{\text{\tiny (M2)}}|_{\cC\cap S}\,.
\eeq
This is the M2 counterpart of \cite[Eq.~(3.25)]{Martucci:2011dn} for F1 strings.   
On the other hand
\beq
\begin{aligned}
-\calp^\s{M} K_\s{M}\d^2\sigma&=\sqrt{-h}\,h^{t\alpha}\del_\alpha X^\s{M} e^\s{A}_\s{M} K_\s{A}\d^2\sigma=\sqrt{-h}\,h^{t\alpha}{\mathfrak e}^{\underline\alpha}_\alpha K_{\underline\alpha}\d^2\sigma\\
&= \sqrt{-h}\,{\mathfrak e}^t_{\underline\alpha}K^{\underline\alpha}\,\d^2\sigma= K^{\underline 0}\,\d{\rm vol}_{\cC\cap S}\,,
\end{aligned}
\eeq
where in the last step we have used ${\mathfrak e}^t_{\underline\alpha}=\delta^{\underline{0}}_{\underline\alpha}\ee^{-D}$ and
\beq
\sqrt{-h}\,\d t\wedge \d^2\sigma={\mathfrak e}^{\underline 0}\wedge{\mathfrak e}^{\underline 1}\wedge {\mathfrak e}^{\underline 2}=\ee^D\sqrt{\hat h}\,\d t\wedge \d^2\sigma = \ee^D\d t\wedge \d{\rm vol}_{\cC\cap S}\,.
\eeq
Hence \eqref{ineq2} is equivalent to
\beq
K^{\underline 0}\,\d{\rm vol}_{\cC\cap S}\geq \Omega^{\text{\tiny (M2)}}|_{\cC\cap S}\,,
\eeq
as stated in \eqref{M2bound}. Furthermore, supersymmetry is preserved if and only if the bound is saturated. 

%%%%%%%%%%%%%%%%%%%%%%%%%%%%%%%%%%%%%%%%%%%%%%%%%%%%%%%%%%%%%%%%%%%%%

% subsection m (end)

\section{Details on type II} % (fold)
\label{app:ii}

In this appendix we describe the computation leading from (\ref{eq:nablaE-ii-0}) to (\ref{eq:nablaE-ii}). We will be less detailed than in section \ref{app:m}; we advise the reader to read that first.

The first step is subtracting the terms on the first line of (\ref{eq:nablaE-ii}); this is lengthy and tedious, but in principle straightforward. Useful identities for this include: 
\begin{equation}
	\Gamma_\s{M} H \Gamma^H = -4 H \, ,\qquad \Gamma_\s{M} H \Gamma^\s{MN} = -4 H \Gamma^\s{N} - \Gamma^\s{N} H\,,
\end{equation}
which follow from (\ref{eq:gag}) and from $\Gamma^\s{M} \Gamma^\s{N} = \Gamma^\s{MN}+ g^\s{MN}1$; the usual definition $\{\Gamma^\s{M},\Gamma^\s{N}\}= 2 g^\s{MN}1$ of Clifford algebra; and (\ref{eq:Gepsbar}). The result is
\begin{equation}\label{eq:nablaE-ii-app}
	\begin{split}
		\ee^{2 \phi} \nabla_\s{M} E^\s{MN} &= \overline{\left({\mathcal D}_\s{M} - \frac18 \Gamma_\s{M} {\mathcal O}\right)\epsilon}\,\Gamma^\s{MPN} \left({\mathcal D}_\s{P} - \frac18 \Gamma_\s{P} {\mathcal O}\right)\epsilon - \frac18 \overline{{\mathcal O} \epsilon} \,\Gamma^\s{N} {\mathcal O } \epsilon\\
		&-\bar \epsilon\, \Gamma^\s{MNP}(D_\s{M} D_\s{P} + [D_\s{M},\Gamma^\s{MNP} {\mathcal A}_\s{P}]) \epsilon- \bar\epsilon\,\left[\sum_{\alpha=1}^6 Q^\s{N}_\alpha\right] \epsilon
	\end{split}
\end{equation}
where the $Q_\alpha^N$ are terms quadratic in the fields, which we will write below; they will need further, less trivial processing. 

\subsection{Linear terms} % (fold)
\label{sub:lin}

We start from the linear terms in (\ref{eq:nablaE-ii-app}). 

The $\bar \epsilon \Gamma^\s{MNP}D_\s{M} D_\s{P} \epsilon$ is evaluated with (\ref{eq:GMN}). For the remaining terms, we recall (\ref{eq:E2-ii-exp}) and (\ref{eq:Dnabla}):
\begin{equation}\label{eq:lin-ii}
	[D_\s{M},\Gamma^\s{MNP} {\mathcal A}_\s{P}]= +\frac14 \Gamma^\s{[M} \nabla_\s{M} H \Gamma^\s{N]} \otimes \sigma_3 - \Gamma^\s{MN} \nabla_\s{M}\dd \phi -2 \Gamma^\s{[M} \nabla_\s{M} {\mathcal F}\, \Gamma^\s{N]}\,.
\end{equation}

The dilaton term gives 
\begin{equation}\label{eq:lin-dil}
	-\bar \epsilon \,\Gamma^\s{MN} \nabla_\s{M}\dd \phi \,\epsilon= 
	- \bar \epsilon (\Gamma^\s{MNP}+2 \Gamma^\s{[M} g^\s{N]P}) \nabla_\s{M} \partial_\s{P} \phi \epsilon = - (\nabla^\s{P} \nabla^\s{N} - g^\s{NP} \nabla^2) \phi \,\bar \epsilon \Gamma_\s{P} \epsilon\,.
\end{equation}
The $H$ term simplifies similar to (\ref{eq:lin}):
\begin{equation}
\begin{split}
	&\frac14 \bar \epsilon\, \Gamma^\s{[M} \nabla_\s{M} H \Gamma^\s{N]} \otimes \sigma_3 \epsilon = -\frac14 \bar \epsilon [(\dd x \wedge + \iota)(\dd x \wedge - \iota)]^\s{[MN]} \nabla_\s{M} H \otimes  \sigma_3 \epsilon\\
	&=\frac14 \bar \epsilon (\dd x^\s{N} \wedge \dd H -\iota^\s{N} * \dd * H) \otimes  \sigma_3 \epsilon\,.
\end{split}	
\end{equation}
Following the same steps, recalling the definition of ${\mathcal F}$ in (\ref{eq:susy-op-ii}) and the self-duality property (\ref{eq:sF}), the RR term becomes
\begin{equation}
	-2 \bar \epsilon \Gamma^\s{[M} \nabla_\s{M} {\mathcal F}\, \Gamma^\s{N]} \epsilon
	= \pm \frac14 \bar \epsilon_1 (\dd x^\s{N} \wedge \dd (\ee^\phi F)) \epsilon_2 +\frac14 \bar \epsilon_2 (\dd x^\s{N} \wedge \dd (\ee^\phi \lambda F)) \epsilon_1\,.
\end{equation}
By (\ref{eq:Gepsbar}), the two terms on the right-hand side are in fact equal to each other.

% subsection lin (end)

\subsection{Quadratic tems} % (fold)
\label{ssub:q}

We now turn to the $Q_\alpha^N$ in (\ref{eq:nablaE-ii-app}). We organized them depending on what fields they contain. 

We first give some useful identities. One is obtained by applying the Fierz identities to the bispinors $\epsilon_a \otimes  \bar\epsilon_a$. As in (\ref{eq:Qfierz}), the only non-zero terms in the sum are those for $k=1$, 5 and 9. Moreover the last is dual to the first. This results in 
\begin{equation}\label{eq:ee-fierz}
\begin{split}
	&32\epsilon_1 \otimes \bar\epsilon_1 = (\bar \epsilon_1 \,\Gamma_\s{M} \epsilon_1) (1+ \Gamma) \Gamma^\s{M} + \frac1{5!} (\bar \epsilon_1\, \Gamma_\s{M_5 \cdots M_1} \epsilon_1) \Gamma^\s{M_1 \cdots M_5}\,,\\
	&32\epsilon_2 \otimes \bar\epsilon_2 = (\bar \epsilon_2 \,\Gamma_\s{M} \epsilon_2) (1\mp \Gamma) \Gamma^\s{M} + \frac1{5!} (\bar \epsilon_2\, \Gamma_\s{M_5 \cdots M_1} \epsilon_2) \Gamma^\s{M_1 \cdots M_5}\,.
\end{split}
\end{equation}

The first is purely quadratic in $H$:
\begin{align}
	64 \bar \epsilon\,Q^\s{N}_1 \epsilon &\equiv  \bar \epsilon\, \Gamma_\s{M} H \Gamma^\s{N} H \Gamma^\s{M} \epsilon = 
	\mathrm{Tr}(\Gamma_\s{M} H \Gamma^\s{N} H \Gamma^\s{M} (\epsilon_1 \otimes \bar\epsilon_1 + \epsilon_2 \otimes \bar\epsilon_2)))\\
	\nonumber
	&\buildrel{(\ref{eq:ee-fierz}),(\ref{eq:gag})}\over=-\frac14\bar \epsilon_1 \Gamma_\s{M} \epsilon_1 \mathrm{Tr}(H \Gamma^\s{N} H  (1-\Gamma) \Gamma^\s{M}) 
	-\frac14\bar \epsilon_2 \Gamma_\s{M} \epsilon_2 \mathrm{Tr}(H \Gamma^\s{N} H  (1\pm\Gamma) \Gamma^\s{M}) \\
	\nonumber
	&\buildrel{(\ref{eq:ii-tr})}\over=16 T^\s{MN}_\s{(H)} \bar \epsilon\,\Gamma_\s{M} \epsilon\,.
\end{align}
With similar steps,
\begin{align}
	\bar \epsilon\,Q^\s{N}_2 \epsilon&\equiv \bar \epsilon\,(\slashed{\partial} \phi \Gamma^\s{N} \slashed{\partial} \phi -2 \nabla^\s{N} \phi \slashed{\partial} \phi) \epsilon = -g^\s{MN}|\dd \phi|^2  \bar \epsilon\,\Gamma_\s{M} \epsilon\,;\\
	\label{eq:Q3}\bar \epsilon\,Q^\s{N}_3 \epsilon &\equiv \bar \epsilon \, \Gamma_\s{M} {\mathcal F} \Gamma^\s{N} {\mathcal F} \Gamma^\s{M} \epsilon \\
	\nonumber&=\frac{\ee^{2 \phi}}{8}\left((\bar \epsilon_1 \, \Gamma_\s{M} \epsilon_1)(T^\s{MN}_\s{F} -*(F \wedge \lambda F)^\s{MN}) 
	+ (\bar \epsilon_2 \, \Gamma_\s{M} \epsilon_2)(T^\s{MN}_\s{F} +*(F \wedge \lambda F))^\s{MN}\right) \,.
\end{align}
In (\ref{eq:Q3}) we have also used (\ref{eq:idx-tr}), and 
\begin{equation}
	(\dd x^\s{M} \wedge \dd x^\s{N} \wedge F \wedge \lambda F)_{10}= - (\iota^\s{M} F)\cdot(\dd x^\s{N} \wedge F)\,,
\end{equation}
which follows from repeated use of (\ref{eq:*dx}) and self-duality (\ref{eq:sF}).

The remaining three $Q^N_\alpha$ require use of (\ref{eq:lrG}). We have
\begin{equation}
\begin{split}
	8 Q^\s{N}_4 &= [\{H,\Gamma^\s{N}\}, \slashed{\partial} \phi]= \partial_\s{M} \phi (\stackrel\leftarrow{\Gamma^\s{M}}\stackrel\leftarrow{\Gamma^\s{N}}- \stackrel\to{\Gamma^\s{M}}\stackrel\to{\Gamma^\s{N}} -2 \stackrel\to{\Gamma^\s{[M}}\stackrel\leftarrow{\Gamma^\s{N]}}) H\\
	&= \partial_\s{M} \phi \left(-(\dd x \wedge - \iota)^2 - (\dd x \wedge + \iota)^2+2 (\dd x\wedge - \iota)(\dd x \wedge + \iota)\right)^\s{[MN]} H\\
	&= -4 \partial_\s{M} \phi \iota^\s{M} \iota^\s{N} H = - 4\iota_{\dd \phi} \iota^\s{N} H 
	= 4 \iota^\s{N} * \dd \phi \wedge * H \,,
\end{split}
\end{equation}
recalling (\ref{eq:*dx}) in the last step. A similar computation, skipping a few steps, gives
\begin{equation}
\begin{split}
	Q^\s{N}_5 &= \slashed{\partial} \phi {\mathcal F} \Gamma^\s{N} - \Gamma^\s{N} {\mathcal F} \slashed{\partial} \phi= 2 \partial_\s{M} \phi \stackrel\to{\Gamma^\s{[M}}\stackrel\leftarrow{\Gamma^\s{N]}} {\mathcal F}\\
	&=\frac{\ee^\phi}8 \left(\mp(1-* \lambda)\dd x^\s{N} \wedge \dd \phi \wedge F \otimes  b^\dagger - (1- \lambda *)\dd x^\s{N} \wedge \dd \phi \wedge \lambda F \otimes  b\right)\,.
\end{split}	
\end{equation}
Finally, the most complicated term is
\begin{align}
	8Q^\s{N}_6 &= [\Gamma_\s{M},H \Gamma^\s{N}] {\mathcal F}\Gamma^\s{M} + \Gamma^\s{M} {\mathcal F}[\Gamma^\s{N} H,\Gamma_\s{M}]\\
	\nonumber&=\mp 8 ( (H \wedge + \iota_H) (\dd x-\iota)^\s{N} + (H \wedge - \iota_H) (\dd x+\iota)^\s{N} ) {\mathcal F}=\mp 16 (H \wedge \dd x^\s{N}- \iota_H \iota^\s{N}) {\mathcal F}\\
	\nonumber & = \mp \ee^\phi (1- *\lambda) H \wedge \dd x^\s{N} \wedge F \otimes  b^\dagger -\ee^\phi (1-\lambda *) H \wedge \dd x^\s{N} \wedge \lambda F \otimes  b\,;
\end{align}	
we defined $\iota_H\equiv \frac1{3!}H_\s{MNP}\iota^\s{M} \iota^\s{N} \iota^\s{P}$, and we used 
\begin{equation}
	\begin{split}
		\stackrel\to{H} + \stackrel\to{\Gamma^\s{M}}\stackrel\leftarrow{H_\s{M}}&= 4 (H \wedge + \iota_H)\,,\\
		\stackrel\leftarrow{H} + \stackrel\leftarrow{\Gamma^\s{M}}\stackrel\to{H_\s{M}}&= 4 (H \wedge - \iota_H)(-1)^{\rm deg}\,,
	\end{split}
\end{equation}
which can in turn be derived from (\ref{eq:lrG}).

Collecting all the results in this subsection finally takes (\ref{eq:nablaE-ii-app}) to (\ref{eq:nablaE-ii}).

% subsection q (end)

% section ii (end)

\section{From M-theory to IIA} % (fold)
\label{sec:m-ii}

Consider the compactification of M-theory on a circle with periodic variable $y\simeq y+1$ with $\ell_{\rm s}=\lp=1$.
Hatted quantities will refer to M-theory. The 11d coordinates are $x^{\,\hat{\!\s{M}}}=(x^\s{M},y)$ and 
the $d=11$ vielbein $\hat e^{\,\hat{\!\s{A}}}=(\hat e^\s{A},\hat e^{\underline 0})$ is related to the 10d one $e^{\s{A}}$ by: 
\begin{equation}
\hat e^\s{A}=\ee^{-\phi/3}e^\s{A}\,,\qquad \hat e^{\underline{10}}\,\ee^{2\phi/3}=\left(\dd y -C_1\right)\,.
\end{equation}
Correspondingly,  $\hat\epsilon\,=\ee^{-\phi/6}\epsilon\equiv\ee^{-\phi/6}(\epsilon_1+\epsilon_2)$ while the 11d gravitino  splits as follows into string frame gravitino and dilatino:
\begin{equation}
\hat\psi=\hat\psi_{\,\hat{\!\s{M}}}\dd x^{\,\hat{\!\s{M}}}=\ee^{-\phi/6}\left(\psi-\frac16\Gamma_{(1)}\lambda\right)+\frac13e^{5\phi/6}\Gamma^{\underline{10}}\lambda\left(\dd y-C_1\right)\,.
\end{equation}

Consider  now a foliation into space-like slices $\hat S=S\times S^1$ and the corresponding asymptotic boundaries $\del\hat S=\del S\times S^1$. By identifying $\Gamma^{\underline{10}}\equiv\Gamma$, the corresponding conserved supercharge reduces as follows
\begin{align}
\hat Q(\hat\epsilon)=&\,-\int_{\del\hat S}\,\overline{\hat\epsilon}\hat\Gamma_{(8)}\wedge\hat\psi\\
\nonumber
=&\, -\int_{\del\hat S}\ee^{-2\phi}\bar\epsilon[e^{-\phi}\Gamma_{(8)}+\Gamma_{(7)}\Gamma\wedge (\dd y -C_1)] \wedge[\psi-\frac16\Gamma_{(1)}\lambda+\frac13e^{\phi}\Gamma\lambda(\dd y-C_1) ]\\
\nonumber
=&\,-\int_{\del\hat S}\ee^{-2\phi}\bar\epsilon \Gamma\Gamma_{(7)}\wedge (\psi-\frac16\Gamma_{(1)}\lambda)\wedge(\dd y -C_1)+\frac13\int_{\hat{\mathcal S}}\ee^{-2\phi}\bar\epsilon\Gamma\Gamma_{(8)}\lambda\wedge(\dd y -C_1)\\
\nonumber
=&\,-\int_{\del\hat S}\ee^{-2\phi}\bar\epsilon \Gamma(\Gamma_{(7)}\wedge \psi-\Gamma_{(8)}\lambda)\wedge(\dd y -C_1)\\
\nonumber
=&\,-\int_{\del S}\ee^{-2\phi}\bar\epsilon \Gamma\left(\Gamma_{(7)}\wedge \psi-\Gamma_{(8)}\lambda\right)\equiv Q(\epsilon)\,.
\end{align}
So we indeed get the type IIA supercharge (\ref{typeIIQ}).

% section m_ii (end)

\bibliography{at,references,sg}
\bibliographystyle{JHEP}

\end{document}